\documentclass{elsart}
\usepackage{graphicx}
\usepackage{amssymb}
\begin{document}

\begin{frontmatter}
\title{Phase-shift analysis of low-energy $\pi^{\pm}p$ elastic-scattering data}
\author[EM]{E. Matsinos{$^*$}},
\author[WS]{W.S. Woolcock},
\author[GC]{G.C. Oades},
\author[GR]{G. Rasche},
\author[AG]{A. Gashi}
\address[EM]{Varian Medical Systems Imaging Laboratory GmbH, T\"{a}fernstrasse 7, CH-5405 Baden-D\"{a}ttwil, Switzerland}
\address[WS]{Department of Theoretical Physics, IAS, The Australian National University, Canberra, ACT 0200, Australia}
\address[GC]{Institute of Physics and Astronomy, Aarhus University, DK-8000 Aarhus C, Denmark}
\address[GR]{Institut f\"{u}r Theoretische Physik der Universit\"{a}t, Winterthurerstrasse 190, CH-8057 Z\"{u}rich, Switzerland}
\address[AG]{Mediscope AG, Alfred Escher-Str. 27, CH-8002 Z\"{u}rich, Switzerland}

\begin{abstract}
Using electromagnetic corrections previously calculated by means of a potential model, we have made a phase-shift analysis of the $\pi^\pm p$ 
elastic-scattering data up to a pion laboratory kinetic energy of $100$ MeV. The hadronic interaction was assumed to be isospin invariant. We found 
that it was possible to obtain self-consistent databases by removing very few measurements. A pion-nucleon model, based on s- and u-channel diagrams 
with $N$ and $\Delta$ in the intermediate states, and $\sigma$ and $\rho$ t-channel exchanges, was f\mbox{}itted to the elastic-scattering database 
obtained after the removal of the outliers. The model-parameter values showed an impressive stability when the database was subjected to dif\mbox{}ferent 
criteria for the rejection of experiments. Our result for the pseudovector $\pi N N$ coupling constant (in the standard form) is $0.0733 \pm 0.0014$. The 
six hadronic phase shifts up to $100$ MeV are given in tabulated form. We also give the values of the $s$-wave scattering lengths and the $p$-wave 
scattering volumes. Big dif\mbox{}ferences in the $s$-wave part of the interaction were observed when comparing our hadronic phase shifts with those of 
the current GWU solution. We demonstrate that the hadronic phase shifts obtained from the analysis of the elastic-scattering data cannot reproduce the 
measurements of the $\pi^- p$ charge-exchange reaction, thus corroborating past evidence that the hadronic interaction violates isospin invariance. 
Assuming the validity of the result obtained within the framework of chiral perturbation theory, that the mass dif\mbox{}ference between the $u$- and the 
$d$-quark has only a very small ef\mbox{}fect on the isospin invariance of the purely hadronic interaction, the isospin-invariance violation revealed by 
the data must arise from the fact that we are dealing with a hadronic interaction which still contains residual ef\mbox{}fects of electromagnetic origin.
\end{abstract}
{$^*$}{Corresponding author. E-mail address: evangelos.matsinos@varian.com. Tel.: +41 56 2030460. Fax: +41 56 2030405}
\end{frontmatter}

\section{Introduction}

In two previous papers \cite{gmorw,gmorww}, we presented the results of a new evaluation of the electromagnetic (EM) corrections which have to 
be applied in a phase-shift analysis (PSA) of the $\pi^\pm p$ elastic-scattering data at low energies (pion laboratory kinetic energy $T \leq 100$ MeV) 
in order to extract the hadronic phase shifts. The calculation used relativised Schr{\"o}dinger equations containing the sum of an EM and a hadronic 
potential; the hadronic potential was assumed to be isospin invariant. We gave reasons for accepting the new corrections as more reliable than the ones 
obtained by the NORDITA group \cite{two} in the late 1970s. For the $\pi^+ p$ scattering data, the corrections $C_{0+}^{+}$, $C_{1-}^{+}$ and 
$C_{1+}^{+}$ for the $s$, $p_{1/2}$ and $p_{3/2}$ waves are given in Table 1 of Ref.~\cite{gmorw}. For the $\pi^- p$ scattering data, the corrections 
$C^{1/2}$, $C^{3/2}$ and $\Delta{\phi}$ are listed in Tables 1-3 of Ref.~\cite{gmorww}.

In this paper, we present the results of a PSA of the $\pi^\pm p$ elastic-scattering data for $T\leq 100$ MeV which was performed using these new EM 
corrections. We have imposed two important restrictions on the experimental data.

First, in contrast to the Karlsruhe analyses \cite{kp} and to the modern GWU (formerly, VPI) solutions \cite{abws}, we have chosen to analyse the 
elastic-scattering data separately. There is an important reason for this decision. As pointed out in Refs.~\cite{gmorw,gmorww}, what we called 
hadronic potentials (in those papers) are not the purely hadronic potentials which model the hadronic dynamics in the absence of the EM interaction; 
they contain residual EM ef\mbox{}fects, and we henceforth call them `electromagnetically modif\mbox{}ied' (em-modif\mbox{}ied, for short) hadronic 
potentials. On page $458$ of Ref.~\cite{gmorw} and page $464$ of Ref.~\cite{gmorww}, we explained that we were not attempting to calculate corrections 
which would remove these residual ef\mbox{}fects. The corrections calculated in Refs.~\cite{gmorw,gmorww} lead to an em-modif\mbox{}ied hadronic 
situation, for which there is evidence that isospin invariance is violated \cite{glk,m}. It is known that, provided that this violation is reasonably 
small, it is still possible to analyse the $\pi^- p$ elastic-scattering data in a framework of formal isospin invariance, borrowing the $I=3/2$ hadronic 
phase shifts from an analysis of $\pi^+ p$ scattering data and then obtaining the $I=1/2$ hadronic phase shifts from an analysis of $\pi^- p$ 
elastic-scattering data. In the present paper, we therefore retain the framework of formal isospin invariance of Refs.~\cite{gmorw,gmorww}. This approach 
enables us to directly investigate the violation of isospin invariance by using the output of our PSA to examine the reproduction of the experimental data 
for the charge-exchange (CX) reaction $\pi^- p\rightarrow \pi^0 n$.

The second important restriction concerns the energy limitation $T \leq 100$ MeV. We think that it is important to analyse this body of data separately, 
and \emph{then} to compare the output of the analysis with that of works which use the entire pion-nucleon ($\pi N$) database as input \cite{kp,abws}, 
or with the result for the $s$-wave threshold parameter $a^{cc}$ for $\pi^- p$ elastic scattering deduced from pionic hydrogen \cite{ss,henn}. There is 
now an abundance of data for $T \leq 100$ MeV from experiments carried out at pion factories, and these data alone are suf\mbox{}f\mbox{}icient to 
determine the $s$- and $p$-wave hadronic phase shifts reliably, as well as the low-energy constants characterising the $\pi N$ interaction near threshold. 
If data from higher energies are included in an analysis, and scaling (f\mbox{}loating) of the dif\mbox{}ferential-cross-section (DCS) measurements is 
allowed, there is the possibility of a systematic rescaling of the low-energy data in order to match the behaviour of the partial-wave amplitudes obtained 
from the higher energies, resulting in scale factors for the low-energy experiments whose average is signif\mbox{}icantly dif\mbox{}ferent from the 
expected value of $1$. In our work, it has been verif\mbox{}ied that the scale factors are not energy dependent and that their average values (for the 
two elastic-scattering processes, separately analysed) are close to $1$.

Implementation of a PSA always involves a decision on where to truncate the partial-wave expansion of the scattering amplitudes. For $T\leq 100$ MeV,
it is suf\mbox{}f\mbox{}icient to retain terms up to $l=3$. In this region, the $d$- and $f$-wave hadronic phase shifts are very small and their EM 
corrections negligible. Nevertheless, these phase shifts need to be included; although their presence does not improve the quality of the f\mbox{}it, 
it induces small changes in the output $s$ and $p$ waves. Since in the GWU analysis \cite{abws}, which incorporates dispersion-relation constraints, the 
$d$ and $f$ waves are determined reliably in the region $T > 100$ MeV, we decided to use their solution. As previously mentioned, the sensitivity of the 
output $s$- and $p$-wave hadronic phase shifts to a variation of the $d$ and $f$ waves is small; hence, the uncertainties due to the $d$- and $f$-wave 
hadronic phase shifts are of no importance when compared to the ones associated with the experimental data being f\mbox{}itted.

Before giving a description of the rest of this paper, we comment on our use of EM corrections which are not complete (stage 1 corrections), but leave 
EM ef\mbox{}fects in the hadronic interaction, which require further (stage 2) corrections. Stage 1 corrections, which we have calculated 
nonperturbatively using a potential model, are reliable estimates of the ef\mbox{}fects of the Coulomb interaction and, in the case of $\pi^- p$ 
scattering, of the external mass dif\mbox{}ferences and the $\gamma n$ channel. They are not superseded by f\mbox{}ield theoretical estimates of the 
stage 2 corrections, which take account of diagrams with internal photon lines and of mass dif\mbox{}ferences in intermediate states. The only reliable 
calculation of stage 2 corrections has been made in Ref.~\cite{gilmr} for the parameter $a^{cc}$, using chiral perturbation theory (ChPT).

In Ref.~\cite{nu}, a variant version of ChPT is used to take account of the full ef\mbox{}fect of the EM interaction on the analysis of $\pi N$ scattering 
data. The authors use only a small amount of experimental data, for $T \lesssim 30$ MeV. Numerical values for EM corrections and $s$-wave scattering 
lengths are not given. From the solid curve for the $\pi^- p$ elastic-scattering $s$-wave `phase shift' in Fig.~$5$ of Ref.~\cite{nu}, one can deduce 
a value of about $0.046$ fm for $a^{cc}$; this number increases to about $0.088$ fm, if part of the EM correction is not taken into account (dashed 
curve in their Fig.~$5$). It is hard to assess what one can learn from these two numbers. To start with, a factor of $2$ hardly represents a correction. 
Further, starting from the latest experimental result of Ref.~\cite{henn} on pionic hydrogen, and taking account of the EM correction calculated in 
Ref.~\cite{gilmr}, a value of $0.132$ fm is obtained for $a^{cc}$, with an error of about $0.004$ fm. There is clearly a very big disagreement between 
this result and that of Ref.~\cite{nu}.

For low-energy $\pi N$ physics, the extension of the work of Ref.~\cite{gilmr} to other threshold parameters and ultimately to the calculation of 
stage 2 corrections at nonzero energies is needed. Until such calculations exist, there is no choice but to work with the stage 1 corrections given 
in Refs.~\cite{gmorw,gmorww}.

Section 2 sets out the formalism which establishes the connection between the em-modif\mbox{}ied hadronic phase shifts and the observables which are 
subject to experimentation. This involves a lengthy chain of connections; to avoid constant reference to various sources of equations, we have decided to 
keep the formalism largely self-contained.

Section 3 lists the experiments in the databases for $\pi^{\pm}p$ elastic scattering and discusses the treatment of the statistical and normalisation 
uncertainties for the various experiments. For both $\pi^+ p$ and $\pi^- p$ elastic scattering, the databases consisted mostly of experimental results
for the DCS and the analysing power, measured at a series of angles for energies between $30$ and $100$ MeV. In addition, for $\pi^+ p$, we used 
measurements of partial-total and total (called total-nuclear in the past) cross sections at a number of energies. Our aim in this PSA was to reject as 
few measurements as possible; whole experiments were removed only when it was beyond doubt that their angular distribution had a shape incompatible 
with the bulk of the data. The optimisation method used required, in addition to the statistical errors on the individual data points, the normalisation 
error of each data set. In the case of cross sections, the normalisation error arises from uncertainties in the beam f\mbox{}lux and target thickness, 
and, in the case of analysing powers, from the uncertainty in the degree of polarisation of the target. There are a few experiments where the 
normalisation error has not been properly reported; in order to treat all experiments on an equal basis, it was necessary to assign normalisation errors 
to all such experiments.

In Section 4, we discuss the statistical procedure followed and the way in which the outliers were identif\mbox{}ied and removed from the databases. In 
order to reject the smallest possible amount of experimental information, simple expansions of the $s$- and $p$-wave $K$-matrix elements were assumed (as 
in Ref.~\cite{fm}); these expansions contain three parameters for each $s$-wave and two for each $p$-wave hadronic phase shift, thus making seven 
parameters in all for each value of the total isospin $I$. In order to determine the seven parameters corresponding to the $I=3/2$ hadronic phase shifts, 
the $\pi^+ p$ elastic-scattering data were f\mbox{}itted f\mbox{}irst. The Arndt-Roper formula \cite{ar} was used in the optimisation. We will explain in 
detail how the data sets were tested for bad shape and normalisation. Consistent with our aim of rejecting as few data as possible, a mild $0.27 \%$ was 
adopted as the signif\mbox{}icance level for rejection, instead of the more standard (among statisticians) value of $1 \%$. In the statistical sense, 
$0.27 \%$ corresponds to a $3 \sigma$ ef\mbox{}fect for the normal distribution. It was necessary to remove only two data sets from the $\pi^+ p$ 
database, and to f\mbox{}loat two sets freely (due to their bad normalisation). Two sets could be saved by removing just one point from each set, and a 
third one was saved by removing two points. We had to reject just $24$ degrees of freedom out of a total of $364$. After the completion of the analysis 
of the $\pi^+ p$ database, the $\pi^- p$ elastic-scattering database was analysed, using the $I=3/2$ hadronic phase shifts obtained from the $\pi^+ p$ 
reaction; the seven parameters for the $I=1/2$ hadronic phase shifts were thus obtained. The same tests for bad shape and normalisation were performed. 
In this case, just one data set had to be removed, one set was freely f\mbox{}loated, and a single point had to be removed from each of two data sets. 
There were just $8$ rejected degrees of freedom out of a total of $336$. After the completion of the two seven-parameter f\mbox{}its and the removal of 
the outliers, we combined the two truncated elastic-scattering databases, to form a single database to be used in the rest of our work.

In Section 5, we make use of a low-energy $\pi N$ model which was described in Ref.~\cite{glmbg} and developed further in Ref.~\cite{m}. This model is 
based on $N$ and $\Delta$ s- and u-channel graphs, and $\sigma$ and $\rho$ t-channel exchanges. The model incorporates the important constraints imposed 
by crossing symmetry. There are now just seven adjustable parameters for f\mbox{}itting the combined truncated elastic-scattering database ($340$ degrees 
of freedom in $\pi^+ p$, $328$ in $\pi^- p$). The f\mbox{}it using the model results in $\chi^2/\mathrm{NDF}=1.308$ (NDF stands for the number of degrees 
of freedom), compared to $1.214$ for the $14$-parameter f\mbox{}it. One has to remark that even the latter f\mbox{}it is poor by conventional 
statistical standards. However, there are two points to make here. One is that the use of the more restrictive $\pi N$ model does not make the f\mbox{}it 
much worse. The second is that, in order to obtain what would usually be considered an acceptable f\mbox{}it, the signif\mbox{}icance level for rejection 
of points would need to be raised to around $10 \%$, with a consequent drastic reduction of the database from $668$ to $562$ entries. The remarkable 
thing, however, is that the output values of the model parameters are very little af\mbox{}fected by this dramatic change in the signif\mbox{}icance 
level. This leads to considerable conf\mbox{}idence in the reliability and stability of the model in f\mbox{}itting the combined truncated 
elastic-scattering database. It seems to us very likely that, for many of the experiments, there has been a substantial underestimation of both the 
statistical and normalisation errors. We therefore think that the best strategy is to work with the low signif\mbox{}icance level of $0.27 \%$, thus 
rejecting as few data as possible, and to include the Birge factor $\sqrt{\chi^2/\mathrm{NDF}}$ in the uncertainties obtained for the model parameters, 
thus adjusting them for the quality of the f\mbox{}it.

In Section 6, we present our results for the em-modif\mbox{}ied hadronic phase shifts (in the form of a table and f\mbox{}igures), as well as for the 
$s$-wave scattering lengths and $p$-wave scattering volumes. We shall compare our results with those from the GWU analysis \cite{abws} and point out 
where the dif\mbox{}ferences lie. We shall also compare the values of the parameter $a^{cc}$ obtained from the scattering lengths given by our PSA and 
from Refs.~\cite{ss,henn}.

In Section 7, we attempt the reproduction of the $\pi^- p$ CX data on the basis of the results obtained in Section 5. The violation of isospin invariance 
at the em-modif\mbox{}ied hadronic level will be demonstrated. Finally, in Section 8, we shall discuss the signif\mbox{}icance of our results and outline 
our understanding of the origin of the isospin-invariance violation in the $\pi N$ system at low energies.

\section{Formalism}

We begin this section by giving, for $\pi^+ p$ elastic scattering, the chain of equations which lead from the em-modif\mbox{}ied hadronic phase shifts 
$\tilde\delta_{l\pm}^{3/2}$ to the measured DCS and analysing power. The use of the symbol $\tilde\delta^{3/2}$ instead of $\delta^h$ (as in 
Ref.~\cite{gmorw}) emphasises that we are dealing with an em-modif\mbox{}ied quantity in a framework of formal isospin invariance. The partial-wave 
amplitudes are def\mbox{}ined as
\begin{equation} \label{eq:pwa}
f^+_{l\pm}=(2iq_c)^{-1}\left\{\exp\left[2i(\tilde{\delta}^{3/2}_{l\pm}+C^+_{l\pm})\right]-1\right\} \, ,
\end{equation}
$q_c$ being the centre-of-mass (CM) momentum of the $\pi^+ p$ system. The EM corrections $C^+_{l\pm}$ are given in Table 1 of Ref.~\cite{gmorw}, for 
$0+$, $1-$ and $1+$, and for $5$ MeV intervals (in $T$) from $10$ to $100$ MeV. The corrections for $l>1$ are very small for $T\leq 100$ MeV and 
were ignored.

The no-spin-f\mbox{}lip and spin-f\mbox{}lip amplitudes $f^+$ and $g^+$ for $\pi^+ p$ elastic scattering are given by
\begin{eqnarray} \label{eq:fplus}
\lefteqn{ f^+ = f^{pc} + f^{ext}_{1\gamma E} + f^{rel}_{1\gamma E} + f^{vp}+ } \nonumber \\
& & + \sum^{\infty}_{l=0} \left\{ (l+1)e^{2i\Sigma_{l+}} f^+_{l+} + le^{2i\Sigma_{l-}} f^+_{l-} \right\} P_{l}(\cos \theta) \, ,
\end{eqnarray}
\begin{equation} \label{eq:gplus}
g^+ = g^{rel}_{1\gamma E} + i\sum_{l=1}^{\infty} \left\{ e^{2i\Sigma _{l+}} f^+_{l+} - e^{2i\Sigma _{l-}} f^+_{l-} \right\}\sin \theta \, P^{'}_{l}(\cos \theta) \, .
\end{equation}

The EM phase shifts $\Sigma_{l\pm}$ have the form
\begin{equation} \label{eq:sigma}
\Sigma_{l\pm} = (\sigma_l-\sigma_0)+\sigma^{ext}_l+\sigma_{l\pm}^{rel}+\sigma_l^{vp} \, ,
\end{equation}
where
\begin{equation} \label{eq:dsigma}
\sigma_l-\sigma_0=\sum^{l}_{n=1}\arctan\left(\frac{\eta f_c}{n} \right) \, ,
\end{equation}
\begin{equation} \label{eq:eta}
\eta=\alpha m_c/q_c , \, f_c=\frac{W^2-m_p^2-\mu_c^2}{2m_cW} \, .
\end{equation}
The quantity $W$ is the total energy in the CM frame, $\mu_c$ and $m_p$ are the masses of $\pi^{\pm}$ and $p$, and $m_c$ is the reduced 
mass of the $\pi^+ p$ system.

The remaining phase shifts in Eq.~(\ref{eq:sigma}) are given in Eqs.~(21)-(23) of \cite{gmorw}; they are the partial-wave projections of the EM
amplitudes in Eqs.~(\ref{eq:fplus}) and (\ref{eq:gplus}), for which the expressions are
\begin{equation} \label{eq:fpc}
f^{pc}=\frac{2\alpha m_c f_c}{t}\exp \left\{ -i\eta f_c\ln (\sin^2 \frac{\theta}{2} ) \right\} \, ,
\end{equation}
\begin{equation} \label{eq:fext}
f^{ext}_{1 \gamma E}=\frac{2\alpha m_c f_c}{t} \left( F^{\pi}F_1^p-1 \right) \, ,
\end{equation}
\begin{equation} \label{eq:frel}
f^{rel}_{1 \gamma E}=\frac{\alpha}{2W} \left\{ \frac{W+m_p}{E+m_p}F_1^p + 2\left( W-m_p+ \frac{t}{4(E+m_p)} \right) F_2^p \right\}F^{\pi} \, ,
\end{equation}
\begin{equation} \label{eq:grel}
g^{rel}_{1 \gamma E}=\frac{i\alpha}{2W \tan (\frac{1}{2}\theta)} \left\{ \frac{W+m_p}{E+m_p}F_1^p + 2\left( W+ \frac{t}{4(E+m_p)} \right) F_2^p \right\}F^{\pi} \, ,
\end{equation}
\begin{equation} \label{eq:fvp}
f^{vp}=-\frac{\alpha \eta f_c}{3\pi q_c} (1- \cos \theta)^{-1} F(\cos \theta) \, ,
\end{equation}
where 
\[
F(\cos \theta)= -\frac{5}{3}+X+(1-\frac{1}{2}X) \sqrt{1+X} \ln \left\{ \frac{\sqrt{1+X}+1}{\sqrt{1+X}-1} \right\} , \, X=-\frac{4m_e^2}{t}.
\]
Here, $m_e$ is the electron mass, $t=-2q_c^2(1-\cos\theta)$, $\theta$ denotes the CM scattering angle, $F^{\pi}$ and $F_{1,2}^p$ are the pion and 
proton EM form factors, respectively, and $E=\sqrt{m_p^2+q_c^2}$. The form factors used for our PSA were approximated by the dipole forms
\begin{equation} \label{eq:F1p}
F_1^p= (1-t/\Lambda_p^2)^{-2}, \, F_2^p= \frac{\kappa_p}{2m_p}F_1^p, \, F^{\pi}=(1-t/\Lambda_{\pi}^2)^{-2} \, ,
\end{equation}
with $\Lambda_p=805$ MeV and $\Lambda_{\pi}=1040$ MeV. Standard notation is used for the quantities $\alpha$ and $\kappa_p$. The ef\mbox{}fect of the 
form factors is very small, and there is no need to use more sophisticated parameterisations or to change the values of $\Lambda_p$ and $\Lambda_{\pi}$ 
used in Refs.~\cite{gmorw,gmorww}.

The experimental observables (DCS and analysing power) are given in terms of $f^+$ and $g^+$ by Eqs.~(1) and (2) of Ref.~\cite{two}:
\begin{equation} \label{eq:dsigmaplus}
\left( \frac{d\sigma}{d\Omega}\right)^+= Z^+(s,t,\Delta E) \left( |f^+|^2+|g^+|^2 \right) \, ,
\end{equation}
\begin{equation} \label{eq:Aplus}
A^+= \frac{2Re(f^+\overline{g}^+)}{|f^+|^2+|g^+|^2} \, .
\end{equation}
The bar denotes complex conjugation. The factor $Z^+(s,t,\Delta E)$ is associated with the emission of (undetected) soft photons, while $\Delta E$ is 
the energy resolution of the experiment and $s$ is the standard Mandelstam variable ($s=W^2$). A detailed discussion of this factor may be found in the 
appendix of Ref.~\cite{two}.

For the $\pi^- p$ system, in addition to the hadronic phase shifts $\tilde{\delta}_{l\pm}^{3/2}$, the $I=1/2$ hadronic phase shifts 
$\tilde{\delta}_{l\pm}^{1/2}$ are introduced. The partial-wave amplitudes are def\mbox{}ined as
\begin{equation} \label{eq:pwfI}
f^{I}_{l\pm}=(2iq_c)^{-1} \left\{ \exp \left[ 2i( \tilde{\delta}_{l\pm}^{I} + C^I_{l\pm} ) \right]-1 \right\}, \, I=1/2, 3/2 \,\,\, ,
\end{equation}
where the EM corrections $C^{1/2}_{l\pm}$, $C^{3/2}_{l\pm}$ are given in Tables 1-3 of Ref.~\cite{gmorww}, for $l\pm=0+, 1-$ and $1+$, at $5$ MeV 
intervals from $10$ to $100$ MeV. Also included in those tables are the EM corrections $\Delta \phi_{0+}$, $\Delta \phi_{1-}$ and 
$\Delta \phi_{1+}$ to the isospin-invariant mixing angle $\phi_0=\arctan (1/\sqrt{2})$. Denoting the $\pi^- p$ channel by $c$ (and the $\pi^0 n$ 
channel by $0$), the partial-wave amplitudes $f^{cc}_{l\pm}$ for $\pi^- p$ elastic scattering have the form
\begin{eqnarray} \label{eq:pwfcc}
\lefteqn{ f^{cc}_{l\pm}=\cos^2(\phi_0+\Delta \phi)f^{1/2} + \sin^2(\phi_0+\Delta \phi)f^{3/2}\,\,\,- } \nonumber \\
& & - \frac{1}{6iq_c} \left\{
2 \bar{\eta}^1 \exp(2i\tilde{\delta}^{1/2} ) +
\bar{\eta}^3 \exp(2i\tilde{\delta}^{3/2} ) +
\frac{8}{3} \eta^{13} \exp \left[i (\tilde{\delta}^{1/2}+ \tilde{\delta}^{3/2}) \right]
\right\} ,
\end{eqnarray}
where, for convenience, we have omitted the subscript $l\pm$ on the right-hand side. The third term on the right-hand side of Eq.~(\ref{eq:pwfcc}) 
takes account of the presence of the third coupled channel $\gamma n$. The values of $\bar{\eta}^1$, $\bar{\eta}^3$ and $\eta^{13}$ for the partial 
waves with $0+$ and $1+$ are given in Table IV of Ref.~\cite{two}. The values for $1-$ are negligible. We have changed the subscripts in \cite{two} 
to superscripts. The numerical values in \cite{two} were derived from known amplitudes for the reactions $\gamma n \rightarrow \pi^- p, \pi^0 n$ using 
three-channel unitarity.

The equations for the $\pi^- p$ elastic-scattering observables are
\begin{equation} \label{eq:dsigmacc}
\left( \frac{d\sigma}{d\Omega}\right)^{cc}= Z^{cc}(s,t,\Delta E) \left( |f^{cc}|^2+|g^{cc}|^2 \right) \, ,
\end{equation}
\begin{equation} \label{eq:Acc}
A^{cc}= \frac{2Re(f^{cc}\overline{g}^{cc})}{|f^{cc}|^2+|g^{cc}|^2} \, ,
\end{equation}
where the amplitudes $f^{cc}$ and $g^{cc}$ are
\begin{eqnarray} \label{eq:fcc}
\lefteqn { f^{cc} = -\bar{f}^{pc} - f^{ext}_{1\gamma E} - f^{rel}_{1\gamma E} - f^{vp} + } \nonumber \\
& & + \sum^{\infty}_{l=0} \left\{ (l+1)e^{-2i\Sigma_{l+}} f^{cc}_{l+} + le^{-2i\Sigma_{l-}} f^{cc}_{l-} \right\} P_{l}(\cos \theta) \, ,
\end{eqnarray}
\begin{equation} \label{eq:gcc}
g^{cc} = -g^{rel}_{1\gamma E} + i\sum_{l=1}^{\infty} \left\{ e^{-2i\Sigma _{l+}} f^{cc}_{l+} - e^{-2i\Sigma_{l-}} f^{cc}_{l-} \right\}\sin \theta \, P^{'}_{l}(\cos \theta) \, .
\end{equation}
All the quantities on the right-hand sides of Eqs.~(\ref{eq:fcc}) and (\ref{eq:gcc}) have already been def\mbox{}ined. The factor $Z^{cc}$ is related to 
$Z^+$ in the manner specif\mbox{}ied in the appendix of Ref.~\cite{two}.

The partial-wave amplitudes $f^{c0}_{l\pm}$ for the CX reaction $\pi^- p \rightarrow \pi^0 n$ have the form
\begin{eqnarray} \label{eq:pwfc0}
\lefteqn{ f^{c0}_{l\pm}=\sqrt{\frac{q_c}{q_0}} \sin(\phi_0+\Delta \phi)\cos(\phi_0+\Delta \phi)
\left( f^{3/2} -f^{1/2}\right) + } \nonumber \\
& & + \frac{1}{2i} \frac{\sqrt{2}}{3\sqrt{q_cq_0}} \nonumber \\
& & \mbox{} \left\{
\bar{\eta}^1 \exp(2i\tilde{\delta}^{1/2} ) -
\bar{\eta}^3 \exp(2i\tilde{\delta}^{3/2} ) -
\frac{2}{3} \eta^{13} \exp \left[i (\tilde{\delta}^{1/2}+ \tilde{\delta}^{3/2}) \right]
\right\} \, ,
\end{eqnarray}
where $q_0$ is the CM momentum in the $\pi^0 n$ system; the other quantities on the right-hand side of Eq.~(\ref{eq:pwfc0}) have the same meaning as in 
Eq.~(\ref{eq:pwfcc}). Again, we have omitted the subscript $l\pm$ on the right-hand side; the second term takes account of the presence of the 
$\gamma n$ channel. In terms of the amplitudes
\begin{equation} \label{eq:fc0}
f^{c0} = \sum^{\infty}_{l=0} \left\{ (l+1)e^{-i\Sigma_{l+}} f^{c0}_{l+} + le^{-i\Sigma_{l-}} f^{c0}_{l-} \right\} P_{l}(\cos \theta) \, ,
\end{equation}
\begin{equation} \label{eq:gc0}
g^{c0} =i\sum_{l=1}^{\infty} \left\{ e^{-i\Sigma _{l+}} f^{c0}_{l+} - e^{-i\Sigma _{l-}} f^{c0}_{l-} \right\}\sin \theta \, P^{'}_{l}(\cos \theta) \, ,
\end{equation}
the DCS for the $\pi^- p$ CX reaction is
\begin{equation} \label{eq:dsigmac0}
\left( \frac{d\sigma}{d\Omega}\right)^{c0}= Z^{c0}(s,\Delta E) \left( \frac{q_0}{q_c} \right) \left( |f^{c0}|^2+|g^{c0}|^2 \right) \, ,
\end{equation}
while the analysing power is given by an expression analogous to Eqs.~(\ref{eq:Aplus}) and (\ref{eq:Acc}). The factor $Z^{c0}$ in
Eq.~(\ref{eq:dsigmac0}) is given in the appendix of Ref.~\cite{two}.

This completes the formalism for the PSA of experiments on low-energy $\pi^{\pm}p$ scattering. The equations leading from em-modif\mbox{}ied hadronic 
phase shifts $\tilde{\delta}_{l\pm}^{1/2}$ and $\tilde{\delta}_{l\pm}^{3/2}$ to the experimental observables have all been given explicitly. 
Reference needs to be made to \cite{gmorw}, for the expressions for the EM phase shifts $\sigma_l^{ext}$, $\sigma_{l\pm}^{rel}$ and $\sigma_l^{vp}$ 
(Eqs.~(21)-(23)) and for Table 1 containing the EM corrections $C_{0+}^{+}$, $C_{1-}^{+}$ and $C_{1+}^{+}$. The corrections $C^{1/2}$, $C^{3/2}$ and 
$\Delta \phi$ for $0+$, $1-$ and $1+$ are found in Tables 1-3 of \cite{gmorww}. For the factors $Z$, appearing in Eqs.~(\ref{eq:dsigmaplus}), 
(\ref{eq:dsigmacc}) and (\ref{eq:dsigmac0}), reference needs to be made to the appendix of Ref.~\cite{two}, while the quantities $\bar{\eta}^1$, 
$\bar{\eta}^3$ and $\eta^{13}$, appearing in the correction terms in Eqs.~(\ref{eq:pwfcc}) and (\ref{eq:pwfc0}) and taking account of the presence of 
the $\gamma n$ channel, are found in Table IV of the same reference. In fact, the factors $Z$ are of minor interest only; we have mentioned them for 
completeness. Since the energy resolution of experiments is only rarely reported, we made the same decision as everyone else involved in analyses of 
low-energy $\pi N$ data, namely, to put the factors $Z$ to $1$.

\section{The databases for $\pi^{\pm}p$ elastic scattering}

The $\pi^+ p$ database comprises the following measurements: DCS \cite{ega}-\cite{cj}, analysing powers \cite{mes,rw}, partial-total cross sections 
\cite{bjk,ef} and total cross sections \cite{cwbbd,ep}. The AULD79 experiment \cite{ega} gave only statistical errors on the data points and did not 
report an overall normalisation uncertainty; we assigned a normalisation error of $5.95 \%$ based on a least-squares f\mbox{}it to the quoted 
meson-factory normalisation errors for the experiments \cite{bgr}-\cite{cj}. Due to the fact that two dif\mbox{}ferent targets were used, the BRACK88 
data \cite{jtbb} were assumed to comprise two independent experiments performed at the same energy ($66.8$ MeV).

The analysing-power experiment of WIESER96 \cite{rw} used two separate targets with dif\mbox{}ferent degrees of polarisation. We therefore separated the 
data from this experiment into two data sets, one with three points and one with four. For this experiment (as well as for that of SEVIOR89 \cite{mes}), 
a normalisation uncertainty of $7.4 \%$ was assigned; this value represents twice the normalisation error of the recent experiment of PATTERSON02 
\cite{jdp} which measured the $\pi^- p$ analysing power. It is hard to decide what to do with experiments for which the normalisation uncertainties were 
not properly reported. Rather than discard them, we made a rough judgment that they should be included in the databases, and assigned 
normalisation errors which are twice those of comparable modern experiments, to take account of their lack of proper reporting and of the age of the 
experiments. As we note later, the exact assignment does not matter.

The partial-total cross sections of KRISS97 \cite{bjk} were obtained at 13 dif\mbox{}ferent energies from $39.5$ to $99.2$ MeV. At each of these 
energies, the cross section for scattering into all (laboratory) angles exceeding $30^\circ$ was measured. At two energies ($66.3$ and $66.8$ MeV), 
the partial-total cross section for scattering into all angles $\geq 20^\circ$ was also measured, using the same beam and target. We therefore 
separated the data into 11 one-point sets and 2 two-point sets, thus giving $13$ data sets in all. The normalisation error on the data points was 
assumed to be $3 \%$; this number appeared in the f\mbox{}irst report of the experiment. For the very similar FRIEDMAN99 experiment \cite{ef}, there are 
$30^\circ$ data at six energies and, in addition, $20^\circ$ data at three of the energies, obtained with the same beam and target. We thus separated 
the data into three one-point sets and three two-point sets, and assigned a normalisation uncertainty of $6 \%$.

The total cross sections of CARTER71 \cite{cwbbd} and PEDRONI78 \cite{ep} were also included in the analysis, each data point being treated as a 
one-point set, with a total error obtained by combining in quadrature the reported errors with a normalisation uncertainty of $6 \%$, twice the 
corresponding error for the experiment of Ref.~\cite{bjk}. The same remarks apply as for the analysing-power experiments discussed above.

The complete initial $\pi^+ p$ database consisted of $364$ entries; a normalisation uncertainty had to be assigned to $39$ entries in total. The 
smallness of this fraction ensures that the output of the analysis is practically insensitive to the precise values of the normalisation uncertainties 
assigned to these 39 measurements. There were $54$ data sets within the full database, $26$ for the DCS, $3$ for the analysing power and $25$ (all 
one- or two-point sets) for the partial-total and total cross sections.

As already explained in Section 1, for the $\pi^- p$ database we conf\mbox{}ined ourselves to elastic scattering only. The published $\pi^- p$ 
partial-total and total cross sections cannot be used, as they contain a large component from CX scattering; the inclusion of these data in any part of 
the analysis would have cast doubt on any conclusions about the violation of isospin invariance. Our database therefore consisted of measurements of the 
DCS \cite{jsf,jtb}, \cite{uw}-\cite{cj} and \cite{mj}, and of the analysing power \cite{mes}, \cite{jca}-\cite{jdp}. The experiment of JANOUSCH97 
\cite{mj} measured the DCS at a single angle ($175^\circ$ in the CM frame) at f\mbox{}ive energies; the data were treated as f\mbox{}ive one-point 
experiments. The experiment of JORAM95 \cite{cj} was considered to comprise eight separate data sets. Data was taken at f\mbox{}ive energies, but, at the 
energies of $32.7, 45.1$ and $68.6$ MeV, the points at higher angles were obtained using a dif\mbox{}ferent target from the one used for lower angles; the 
data obtained with these dif\mbox{}ferent targets were put in separate sets. Only in the case of the analysing-power experiments ALDER83 \cite{jca} and 
SEVIOR89 \cite{mes} did we have to assign a normalisation error, namely the $7.4 \%$ value which was used in the case of the two $\pi^+ p$ analysing-power 
experiments. Thus, the assignment of a normalisation uncertainty was necessary for only $11$ points out of $336$. In the full database, there were $36$ 
data sets ($27$ for the DCS and $9$ for the analysing power).

After our PSA was completed, new experimental data appeared \cite{meier}; this consists of analysing-power measurements at energies between $45.2$ and 
$87.2$ MeV, with $25$ data points for $\pi^+ p$ and $3$ for $\pi^- p$ elastic scattering. For these measurements, three dif\mbox{}ferent targets were 
used, for each of which there was a determination of the target polarisation; therefore, the measurements must be assigned to just three data sets. In 
each of these sets, the measurements correspond to more than one energy, and, in one case, to both $\pi^+ p$ and $\pi^- p$ scattering. Unfortunately, it 
has not been possible to include these data in our analysis. Their inclusion would have involved substantial modif\mbox{}ications in our database 
structure and analysis software as, at present, our data sets are characterised by a single energy and a single reaction type. Nevertheless, we shall 
show (at the end of Section 6) that the data of Ref.~\cite{meier} are well reproduced by the output of our PSA. We shall also give reasons why the 
inclusion of these measurements would have made a negligible dif\mbox{}ference to our results.

\section{The statistical method and the identif\mbox{}ication of outliers}

The Arndt-Roper formula \cite{ar} was used in the optimisation; the quantity which was minimised is the standard $\chi^2$, including a term which takes 
account of the f\mbox{}loating of each data set. The contribution of the $j^{th}$ data set to the overall $\chi^2$ is
\begin{equation} \label{eq:chijsq}
\chi_j^2=\sum_{i=1}^{N_j} \left\{ \frac{ z_jy_{ij}^{th}-y_{ij}^{exp} }{\delta y_{ij}^{exp} } \right\}^2 + 
\left( \frac{z_j-1}{\delta z_j} \right)^2 \, .
\end{equation}
In Eq.~(\ref{eq:chijsq}), $y_{ij}^{exp}$ denotes the $i^{th}$ data point in the $j^{th}$ data set, $y_{ij}^{th}$ the corresponding f\mbox{}itted value 
(also referred to as theoretical), $\delta y_{ij}^{exp}$ the statistical uncertainty associated with the $y_{ij}^{exp}$ measurement, $z_j$ a 
scale factor applying to the entire data set, $\delta z_j$ the relative normalisation error and $N_j$ the number of points in the set. The 
f\mbox{}itted values $y_{ij}^{th}$ are generated by means of parameterised forms of the em-modif\mbox{}ied $s$- and $p$-wave amplitudes. The values of 
the scale factor $z_j$ are determined for each individual data set in order to minimise the contribution $\chi_j^2$. For each data set, there is a unique 
solution for $z_j$:
\begin{equation} \label{eq:zj}
z_j = \frac{\sum_{i=1}^{N_j} y_{ij}^{th} y_{ij}^{exp} / (\delta y_{ij}^{exp} )^2 + (\delta z_j)^{-2}} 
{\sum_{i=1}^{N_j} (y_{ij}^{th} / \delta y_{ij}^{exp})^2 + (\delta z_j)^{-2}} \, ,
\end{equation}
which leads to the value
\begin{equation} \label{eq:chijsqmin}
\left(\chi_j^2\right)_{min} = \sum_{i=1}^{N_j} \frac{ (y_{ij}^{th}-y_{ij}^{exp})^2}{(\delta y_{ij}^{exp})^2 } -
\frac {\left\{ \sum_{i=1}^{N_j}y_{ij}^{th}(y_{ij}^{th}-y_{ij}^{exp}) / (\delta y_{ij}^{exp} )^2 \right\}^2}
{ \sum_{i=1}^{N_j}(y_{ij}^{th}/\delta y_{ij}^{exp})^2 + (\delta z_j)^{-2} } \, .
\end{equation}
The sum of the values $(\chi_j^2)_{min}$ for all data sets $j$ will be denoted simply by $\chi^2$. This total $\chi^2$ for the whole database is a 
function of the parameters which appear in the parameterisation of the $s$- and $p$-wave amplitudes; these parameters were varied until $\chi^2$ 
attained its minimum value $\chi^2_{min}$.

Note that, for a one-point set $(N_j=1)$, Eq.~(\ref{eq:chijsqmin}) reduces to
\begin{equation} \label{eq:chijsqmin1}
\left(\chi_j^2\right)_{min} = \frac{ (y_{j}^{th}-y_{j}^{exp})^2}{(\delta y_{j}^{exp})^2 + (\delta z_j)^2(y_j^{th})^2 } \, .
\end{equation}
The contribution of a one-point set to the overall $\chi^2$ can therefore be calculated from a total uncertainty obtained by adding in quadrature 
the statistical error $\delta y_j^{exp}$ and the normalisation error $(\delta z_j)|y_j^{th}|$. Eqs.~(\ref{eq:chijsqmin}) and (\ref{eq:chijsqmin1}) were 
used to calculate the values of $(\chi_j^2)_{min}$ for each data set. The values of the scale factors $z_j$ (Eq.~(\ref{eq:zj})) need to be calculated 
only once, at the end of the optimisation.

In order to give the data maximal freedom in the process of identifying the outliers, the two elastic-scattering reactions were analysed separately 
using simple expansions of the $s$- and $p$-wave $K$-matrix elements. For $\pi^+ p$ elastic scattering, the $s$-wave phase shift was 
parameterised as
\begin{equation} \label{eq:delta0+3/2}
q_c \cot \tilde{\delta}_{0+}^{3/2}=(\tilde{a}_{0+}^{3/2})^{-1} +b_3\epsilon
+c_3\epsilon^2 \, ,
\end{equation}
where $\epsilon=\sqrt{\mu_c^2+q_c^2}-\mu_c$, while the $p_{1/2}$-wave phase shift was parameterised according to the form
\begin{equation} \label{eq:delta1-3/2}
\tan \tilde{\delta}_{1-}^{3/2}/q_c=d_{31}\epsilon+e_{31}\epsilon^2 \, .
\end{equation}

Since the $p_{3/2}$ wave contains the $\Delta$($1232$) resonance, a resonant piece in Breit-Wigner form was added to the 
background term, thus leading to the equation
\begin{equation} \label{eq:delta1+3/2}
\tan \tilde{\delta}_{1+}^{3/2}/q_c=d_{33}\epsilon+e_{33}\epsilon^2+ \frac{\Gamma_{\Delta}m_{\Delta}^2}{q_{\Delta}^3W} \frac{q_c^2}{m_{\Delta}^2-W^2} \, ,
\end{equation}
where $q_\Delta$ is the value of $q_c$ at the resonance position. Since in Eq.~(\ref{eq:delta1+3/2}) we are parameterising a real quantity, the phase 
factor which appears in the usual expression for the $\Delta$ amplitude is absent. The third term on the right-hand side of Eq.~(\ref{eq:delta1+3/2}) 
is the standard resonance contribution given in Ref.~\cite{ew}. Since we are assuming a framework of formal isospin invariance for hadronic 
phase shifts and threshold parameters, we took the average value $m_\Delta=1232$ MeV from Ref.~\cite{pdg}, as well as the width $\Gamma_\Delta=120$ MeV. 
As the resonance position is around $T=190$ MeV, the exact resonance parameters are not important; small changes in these parameters will be absorbed by 
$d_{33}$ and $e_{33}$. For the PSA of the $\pi^+ p$ elastic-scattering data, the seven parameters in Eqs.~(\ref{eq:delta0+3/2})-(\ref{eq:delta1+3/2}) 
were varied until $\chi^2$ was minimised. The parameterisation described in Eqs.~(\ref{eq:delta0+3/2})-(\ref{eq:delta1+3/2}) was f\mbox{}irst introduced 
(and successfully applied to $\pi^+ p$ elastic scattering) in Ref.~\cite{fm}.

We pause here to note that, for a data set containing $N_j$ points, there were in fact $(N_j+1)$ measurements made, the extra one relating to the 
absolute normalisation of the experiment. When any f\mbox{}it is made using Eq.~(\ref{eq:chijsq}) for $\chi_j^2$, a penalty is imposed for each of the 
data points and for the deviation of the scale factor $z_j$ from 1. Since the f\mbox{}it involves the f\mbox{}ixing of each $z_j$ at the value given in 
Eq.~(\ref{eq:zj}), the actual number of degrees of freedom associated with the $j^{th}$ data set is just $N_j$. As we shall see in a moment, the shapes 
of the data sets were tested by a method in which the scale factors $z_j$ were varied without penalty (free f\mbox{}loating). Furthermore, certain data 
sets were found to have a bad normalisation and were freely f\mbox{}loated in the f\mbox{}inal f\mbox{}its. In all these cases of free f\mbox{}loating, 
the experimental determination of the normalisation was ignored, with the result that the number of degrees of freedom associated with each such set was 
reduced from $N_j$ to $(N_j-1)$.

The use of the parametric forms in Eqs.~(\ref{eq:delta0+3/2})-(\ref{eq:delta1+3/2}), which do not impose any theoretical constraints except for the 
known threshold behaviour, ensures that the existence of any outliers in the database cannot be attributed to the inability of the parametric forms to 
describe the hadronic phase shifts, but indicates problems with some of the data points. The f\mbox{}irst step was to identify any data sets with a shape 
inconsistent with the bulk of the data. To do this, at the end of each iteration in the optimisation scheme, each data set with $N_j>1$ was f\mbox{}loated 
freely; this means that the second term on the right-hand side of Eq.~(\ref{eq:chijsq}) was omitted and $z_j$ was chosen in such a way as to minimise 
the f\mbox{}irst term. Its minimum value $(\chi^2_j)_{stat}$ is given by Eq.~(\ref{eq:chijsqmin}), with $(\delta z_j)^{-2}$ removed from the denominator 
of the second term on the right-hand side. For each data set, the probability was then calculated that the observed statistical variation 
$(\chi^2_j)_{stat}$ for $(N_j-1)$ degrees of freedom could be attributed to random f\mbox{}luctuations. If this probability was below the chosen 
signif\mbox{}icance level ($\mathrm{p}_{min}=0.0027$), it was necessary to eliminate data points from the set. In some cases, the removal of either one 
or two points with the largest contribution to $(\chi^2_j)_{stat}$ raised the p-value for the remainder of the data set above $\mathrm{p}_{min}$. Such 
points were then removed (one at a time) and the analysis was repeated after each removal. In the case of two data sets (BRACK90 \cite{jtbbb} at $66.8$ 
MeV and JORAM95 \cite{cj} at $32.7$ MeV), the p-value was still below $\mathrm{p}_{min}$ after the removal of two points; these two sets (with $11$ 
points and $7$ points, respectively) were removed from the database. The two removed sets have extremely low p-values and stand out dramatically from the 
rest of the $\pi^+ p$ data. In addition, just four individual points needed to be removed, two from JORAM95 at $44.6$ MeV (at $30.74^\circ$ and 
$35.40^\circ$), one from JORAM95 at $32.2$ MeV (at $37.40^\circ$) and one from JORAM95 at $45.1$ MeV (at $124.42^\circ$). For the overall consistency 
of the analysis, any points with a contribution $>9$ to $(\chi^2_j)_{stat}$ had to be removed (since $\mathrm{p}_{min}$ corresponds to a $3\sigma$ 
ef\mbox{}fect for the normal distribution), even though the data set to which they belonged had a p-value greater than $\mathrm{p}_{min}$. This was the 
reason for the elimination of two of the four points just mentioned.

The second step was to investigate the normalisation of the data sets. The p-values corresponding to the scaling contribution to $(\chi^2_j)_{min}$ 
(the second term in Eq.~(\ref{eq:chijsq}), with $z_j$ set at the value obtained using Eq.~(\ref{eq:zj})) were calculated for all the data sets. The 
experiment with the lowest p-value (well below $\mathrm{p}_{min}$) was BRACK86 \cite{jtb} at $66.8$ MeV; this experiment was freely f\mbox{}loated in the 
subsequent f\mbox{}its. A second data set, BRACK86 at $86.8$ MeV, also needed to be freely f\mbox{}loated after the PSA was repeated. After that, no more 
data sets had to be freely f\mbox{}loated. After the reduction in the number of degrees of freedom by $24$, as described above, we were left with a 
$\pi^+ p$ database comprising $52$ data sets with $340$ degrees of freedom. The surviving data sets and the corresponding numbers of degrees of freedom 
are listed in Table \ref{tab:1-table}.

Since seven parameters were used to generate the f\mbox{}itted values, the number of degrees of freedom for the initial f\mbox{}it to the full database 
was $357$; the minimum value of $\chi^2$ was $673.9$. For the truncated database with $333$ degrees of freedom, the minimum value of $\chi^2$ was $425.2$, 
an impressive decrease by $248.7$ units after eliminating a mere $24$ degrees of freedom. At the same time, the p-value of the f\mbox{}it increased by 
$17$ orders of magnitude. The f\mbox{}it to the truncated database detailed in Table \ref{tab:1-table} corresponds to a p-value of $4.62\cdot 10^{-4}$, 
which may appear to be rather poor. Questions about the quality of the f\mbox{}it and the choice of the criterion for the rejection of data points will be 
further discussed in Section 5.

The $I=3/2$ amplitudes were f\mbox{}ixed from the last f\mbox{}it to the $\pi^+ p$ data, made after all the outliers were removed from the database; they 
were then imported into the analysis of the full $\pi^- p$ database. For this analysis, another seven parameters were introduced to parameterise the $s$- 
and $p$-wave $I=1/2$ components of the scattering amplitude. As for the $\pi^+ p$ case, these parameters were varied in order to f\mbox{}ind the minimum 
of the $\chi^2$ function. The same parametric forms were used as in Eqs.~(\ref{eq:delta0+3/2})-(\ref{eq:delta1+3/2}), with the parameters 
$\tilde{a}_{0+}^{1/2}$, $b_1$, $c_1$, $d_{11}$, $e_{11}$, $d_{13}$, $e_{13}$. Of course, there is no resonance term in the expression for 
$\tilde\delta_{1+}^{1/2}$; instead, it is necessary to add the contribution of the Roper resonance to $\tilde\delta_{1-}^{1/2}$:
\begin{equation} \label{eq:delta1-1/2}
\tan \tilde{\delta}_{1-}^{1/2}/q_c=d_{11}\epsilon+e_{11}\epsilon^2+ \frac{\Gamma_{N}m_{N}^2}{q_{N}^3W} \frac{q_c^2}{m_{N}^2-W^2} \, ,
\end{equation}
with $m_N=1440$ MeV, $\Gamma_N=227.5$ MeV and $q_N$ denoting the CM momentum at the Roper-resonance position. As we are dealing with energies below the 
pion-production threshold, the value of $\Gamma_N$ is the elastic width. Inspection of the values of $\Gamma_N$, which are considered in the evaluation 
performed by the Particle Data Group (PDG) \cite{pdg}, reveals a rather disturbing variation; the numbers quoted there (page 868) range between $135$ 
and $545$ MeV. According to the PDG, the best estimate of the $\Gamma_N$ value lies around the centre of this interval; in our work, we used their 
recommendation. In fact, the exact value used for $\Gamma_N$ is of no consequence for our purpose. The Roper resonance makes only a very small 
contribution, even near $100$ MeV, and any change in $\Gamma_N$ will be compensated by changes in the parameters $d_{11}$ and $e_{11}$.

The $\pi^- p$ database was subjected to the same tests of the shape and normalisation of the data sets. In this case, only one data set (BRACK90 
\cite{jtbbb} at $66.8$ MeV, with f\mbox{}ive data points) needed to be removed. Additionally, two single points (the $36.70^\circ$ point of BRACK95 
\cite{jtbbbb} at $98.1$ MeV and the $15.55^\circ$ point of WIEDNER89 \cite{uw} at $54.3$ MeV) had to be rejected. The only data set which needed to be 
freely f\mbox{}loated because of poor normalisation was that of WIEDNER89. The truncated database for $\pi^- p$ elastic scattering f\mbox{}inally 
consisted of $35$ data sets with $328$ degrees of freedom (see Table \ref{tab:2-table}). The number of points in the full $\pi^- p$ elastic-scattering 
database was $336$ and the minimum value of $\chi^2$ was $531.9$ for $329$ degrees of freedom. After the removal of the outliers, there was a spectacular 
drop of $\chi^2_{min}$ to $373.3$ for $321$ degrees of freedom. The p-value for the f\mbox{}it increased by $9$ orders of magnitude, to a value of 
$2.34\cdot 10^{-2}$, indicating a fairly good f\mbox{}it. Interestingly, Table \ref{tab:2-table} shows that the p-values for all the data sets are above 
$0.01$. Therefore, increasing $\mathrm{p}_{min}$ from $0.0027$ to $0.01$ would not af\mbox{}fect the $\pi^- p$ elastic-scattering database.

After all the outliers were removed, a f\mbox{}it to the combined truncated elastic-scattering databases (detailed in Tables \ref{tab:1-table} and 
\ref{tab:2-table}) was made, using $14$ parameters, in order to examine whether any additional points (or even data sets) had to be removed. None were 
identif\mbox{}ied, thus satisfying us that the two truncated elastic-scattering databases are self-consistent and can be used as the starting point for 
further analysis. Before proceeding further, three issues need to be addressed.

First, judged solely on the basis of p-values, it appears that the two truncated databases (that is, $\pi^+ p$ and $\pi^- p$) are not of the same 
quality. However, there is a proper statistical measure of such a comparison between two quantities following the $\chi^2$ distribution. In order to 
prove that the two databases are of dif\mbox{}ferent quality (that is, that they have not been sampled from the same distribution), the ratio
\begin{equation} \label{eq:grossF}
F=\frac{\chi_+^2/N\!D\!F\!_+}{\chi_-^2/N\!D\!F\!_-}
\end{equation}
has to be signif\mbox{}icantly dif\mbox{}ferent from $1$. In this formula, the subscripts $+$ and $-$ denote the two scattering reactions. The ratio $F$ 
follows Fisher's ($F$) distribution. Using the numbers which come from our optimal f\mbox{}its to the two databases separately, we obtain for $F$ the 
value of $1.098$ for $N\!D\!F\!_+=333$ and $N\!D\!F\!_-=321$ degrees of freedom. However, the lowest value of $F$ which would demonstrate a statistically 
signif\mbox{}icant ef\mbox{}fect (at the $95 \%$ conf\mbox{}idence level) for the given degrees of freedom is $1.2$, well above the value $1.098$. The 
p-value corresponding to this value of $F$ ($1.098$) is about $0.2$, far too large to indicate any signif\mbox{}icance. Thus, there is no evidence that 
our two truncated elastic-scattering databases are of dif\mbox{}ferent quality.

The second issue relates to the distribution of the residuals as they come out of the separate f\mbox{}its to the two databases. This is an issue which 
has to be carefully investigated in any optimisation procedure. For instance, pathological cases may result in asymmetrical distributions, usually created 
by large numbers of outliers or by the inability of the parametric model to account for the input measurements. If this is the case, the optimal values of 
the parameters obtained from the f\mbox{}it are bound to be wrong; fortunately, this is not the case in our f\mbox{}its. If, on the other hand, the 
distribution of the residuals turns out to be symmetrical, it has to be investigated whether it corresponds to the form of the quantity chosen for the 
minimisation; this is a necessary condition for the self-consistency of the optimisation scheme. The choice of $\chi^2$ must yield normal distributions 
of the residuals. For each of the databases, we found that the residuals were indeed distributed normally, with a mean of $0$ and almost identical 
standard deviations. It is important to understand one subtle point relating to the distribution of the residuals in low-energy $\pi N$ elastic 
scattering. In Ref.~\cite{m}, this distribution was found to be lorentzian; in the present work, the distribution of the residuals was normal. One might 
then erroneously conclude that the two analyses are in conf\mbox{}lict; in fact, there is none. The point is that no f\mbox{}loating of the data sets was 
allowed in Ref.~\cite{m}; for each data point, the normalisation uncertainty was combined in quadrature with the statistical one to yield an overall 
error. Had we followed the same strategy in the present work, we would also have found lorentzian distributions. However, the f\mbox{}loating introduced 
in Eq.~(\ref{eq:chijsq}) transforms the lorentzian distributions into normal; minimising the quantity def\mbox{}ined in Eq.~(\ref{eq:chijsq}) leads to 
better clustering of the residuals because each data set is allowed to f\mbox{}loat as a whole.

The f\mbox{}inal remark concerns the scale factors $z_j$ obtained in the f\mbox{}its, as listed in Tables \ref{tab:1-table} and \ref{tab:2-table}. For a 
satisfactory f\mbox{}it, the data sets which have to be scaled `upwards' should (more or less) be balanced by the ones which have to be scaled 
`downwards'. Additionally, the energy dependence of the scale factors over the energy range of the analysis should not be signif\mbox{}icant. If these 
prerequisites are not fulf\mbox{}illed, the parametric forms used in the f\mbox{}its cannot adequately reproduce the data over the entire energy range 
and biases are introduced into the analysis. For both the $\pi^+ p$ and $\pi^- p$ DCS data, the values of $z_j$ which lie above and below $1$ roughly 
balance each other and there is no discernible energy dependence. There are too few analysing-power data to make a statement. For the $\pi^+ p$ 
partial-total and total cross sections, all the scale factors are above $1$; however, most of them cluster very close to $1$ and there is no 
signif\mbox{}icant energy dependence. It is interesting to note that when the p-values of the data sets listed in Tables \ref{tab:1-table} and 
\ref{tab:2-table} are plotted as a function of the energy, for $\pi^+ p$ and $\pi^- p$ separately, in neither case is there any evidence of a systematic 
behaviour. In other words, there is no subrange of the full $30$ to $100$ MeV energy range for which the data is better or worse f\mbox{}itted than for 
the rest of the range. We believe that it is essential for any PSA, over whatever energy range, to address the issues we have just considered in this 
section. Having satisf\mbox{}ied ourselves that we have self-consistent $\pi^+ p$ and $\pi^- p$ elastic-scattering databases, no bias in the scale 
factors and an appropriate quantity to be minimised, we can proceed to analyse the combined truncated elastic-scattering database in the framework of the 
$\pi N$ model of Ref.~\cite{glmbg}.

\section{Model parameterisation of the hadronic phase shifts}

\subsection{$\pi^+ p$ and $\pi^- p$ overall weights}

Since the analysis assumes a framework of formal isospin invariance for the hadronic phase shifts and threshold parameters, it is necessary to give the 
two elastic-scattering reactions equal weight. This was achieved by multiplying $(\chi^2_j)_{min}$ (see Eq.~(\ref{eq:chijsq})) for each $\pi^+ p$ data 
set by
\[
w_+=\frac{N\!_+ + N\!_-}{2N\!_+}
\]
and for each $\pi^- p$ data set by
\[
w_-=\frac{N\!_+ + N\!_-}{2N\!_-} \, ,
\]
with $N\!_+=340$ and $N\!_-=328$; we then added these quantities for all the data sets to obtain the overall $\chi^2$ value. The application of the 
overall weights for the two reactions was made as a matter of principle; its ef\mbox{}fect on the PSA is very small.

\subsection{The $\pi N$ model}

To extract the hadronic component of the $\pi N$ interaction from experimental data, it is necessary to introduce a way to model the interaction. So 
far in this paper, we have used expansions of the hadronic phase shifts in terms of the energy. In a moment, we will use a model based on Feynman 
diagrams. Whatever the model, one must then introduce the EM ef\mbox{}fects (as contributions to the hadronic phase shifts and partial-wave amplitudes) 
and use an optimisation procedure in which the model parameters are varied to achieve the best f\mbox{}it to the data. Expansions of the hadronic phase 
shifts in terms of the energy, taking unitarity into account by using the $K$-matrix formalism, are general, but cannot provide any insight into the 
physical processes involved. On the other hand, the use of Feynman diagrams involves a choice of the diagrams to be included in the model, but is able to 
yield an understanding of the dynamics of the $\pi N$ interaction.

Models based on Feynman diagrams suf\mbox{}fer from two problems. First, while the diagrams are chosen to include the contributions of the 
signif\mbox{}icant processes involved in the interaction, there will always be small contributions which are omitted and are absorbed into the dominant 
ones. Second, since the em-modif\mbox{}ied hadronic amplitudes are being approximated by forms which respect formal isospin invariance, it is necessary 
to assign single masses to each of the hadronic multiplets appearing in the model (in our case, $\pi$, $N$, $\Delta$ and $\rho$). All this means that the 
model parameters (coupling constants and vertex factors) become ef\mbox{}fective, and that the amplitudes calculated from such models contain residual 
ef\mbox{}fects of EM origin. The former is a reminder that the errors on the parameters derived from a f\mbox{}it to $\pi N$ data will be underestimated. 
For the latter, we have already pointed out in Section 1 that the hadronic phase shifts being modelled are modif\mbox{}ied by EM ef\mbox{}fects which are 
not included in the calculation of the EM corrections given in Refs.~\cite{gmorw,gmorww}.

In our evaluation of the EM corrections and in the analysis of the $\pi^\pm p$ elastic-scattering data using a $\pi N$ model, we have remained 
self-consistent by f\mbox{}ixing the hadronic masses of pions and nucleons at $\mu_c$ and $m_p$, respectively. These are not the true hadronic masses. 
The hadronic masses of the pions and nucleons are discussed in Ref.~\cite{nu}. The hadronic masses of $\pi^\pm$ and $\pi^0$ are almost identical, 
and dif\mbox{}fer by at most $1$ MeV from the physical mass $\mu_0$ of $\pi^0$. The hadronic mass of the neutron is about $2$ MeV greater than that of 
the proton, as a result of the dif\mbox{}ference between the masses of the $u$- and the $d$-quark. It is noteworthy, however, that the work of 
Ref.~\cite{gilmr} concludes that, despite this mass dif\mbox{}ference, ChPT for the hadronic $\pi N$ interaction is to a good approximation isospin 
invariant, and characterised by single hadronic masses for the pions and nucleons. Ref.~\cite{gilmr} makes the conventional choice of these hadronic 
masses as $\mu_c$ and $m_p$. This choice has been made in previous studies of the low-energy $\pi N$ system \cite{gmorw}-\cite{abws} and we have retained 
it in the present work. This standard choice envisages what might be called a `partially hadronic world', in which the $\pi N$ interaction is isospin 
invariant, but the pion and nucleon masses are $\mu_c$ and $m_p$, respectively. It therefore needs to be borne in mind that the model-derived 
hadronic phase shifts given in the rest of this paper are not true hadronic quantities, but contain residual EM contributions, due to the 
incompleteness of the EM corrections and to the dif\mbox{}ference between the chosen hadronic masses and the true ones.

Since for the analysis of the $\pi^\pm p$ elastic-scattering data we are using a framework of formal isospin invariance, the em-modif\mbox{}ied hadronic 
interaction has been modelled by using the parameterisation of Ref.~\cite{glmbg}. This model is isospin invariant and incorporates the important 
constraints of crossing symmetry and unitarity. The ability of the model to account for the bulk of the elastic-scattering data at least up to the 
$\Delta$ resonance has been convincingly demonstrated. The main diagrams on which the model is based are graphs with scalar-isoscalar ($I=J=0$) and 
vector-isovector ($I=J=1$) t-channel exchanges, as well as the $N$ and $\Delta$ s- and u-channel graphs. The main contributions to the partial-wave 
amplitudes from these diagrams have been given in detail in Ref.~\cite{glmbg}. The small contributions from the six well-established four-star $s$ and 
$p$ higher baryon resonances with masses up to $2$ GeV were also included in the model; in fact, the only such state with a signif\mbox{}icant 
contribution is the Roper resonance. The tensor component of the $I=J=0$ t-channel exchange was added in Ref.~\cite{m}.

The $I=J=0$ t-channel contribution to the amplitudes is approximated in the model by a broad $\pi \pi$ resonance, characterised by two parameters, 
$G_\sigma$ and $K_\sigma$. Its exact position has practically no ef\mbox{}fect on the description of the $\pi N$ scattering data or on the f\mbox{}itted 
values of $G_\sigma$ and $K_\sigma$, and for a long time has been f\mbox{}ixed at $860$ MeV. The $I=J=1$ t-channel contribution is described by the 
$\rho$-meson, with $m_\rho=770$ MeV, which introduces two additional parameters, $G_\rho$ and $K_\rho$. The contributions of the s- and u-channel graphs 
with an intermediate $N$ involve the $\pi NN$ coupling constant $g_{\pi NN}$ and one further parameter $x$ which represents the pseudoscalar admixture in 
the $\pi NN$ vertex; for pure pseudovector coupling, $x=0$. Finally, the contributions of the graphs with an intermediate $\Delta$ introduce the coupling 
constant $g_{\pi N \Delta}$ and one additional parameter $Z$ associated with the spin-$1/2$ admixture in the $\Delta$ f\mbox{}ield. The higher baryon 
resonances do not introduce any parameters.

When a f\mbox{}it to the data is made using all the eight parameters just described, it turns out that there is a strong correlation between $G_\sigma$, 
$G_\rho$ and $x$, which makes it impossible to determine the values of all three quantities. One of the options available is to set $x$ to $0$; this 
choice is usually adopted in ef\mbox{}fective f\mbox{}ield-theoretical models of low-energy $\pi N$ scattering. Thus, seven parameters were used in the 
f\mbox{}it to the combined truncated elastic-scattering database: $G_\sigma$, $K_\sigma$, $G_\rho$, $K_\rho$, $g_{\pi NN}$, $g_{\pi N \Delta}$ and $Z$.

\subsection{Fits and results}

As described in Section 4, the choice of the probability value below which points are removed from the databases is dif\mbox{}f\mbox{}icult and 
subjective. In order to reject as few points as possible, we adopted a very low value of $\mathrm{p}_{min}$ ($0.0027$, the value associated with 
$3\sigma$ ef\mbox{}fects). Recognising the arbitrariness in the choice of $\mathrm{p}_{min}$, we consider that, in order to have conf\mbox{}idence 
in the reliability of our analysis, it is necessary to check that the f\mbox{}itted values of the seven model parameters remain stable over a 
reasonably broad range of $\mathrm{p}_{min}$ values. Thus, in addition to $0.0027$, the analysis was performed with a database reduced by using 
$\mathrm{p}_{min}$-value cuts of $0.01$, $0.05$ and $0.10$. The value of $0.05$ is close to a $2\sigma$ ef\mbox{}fect; points rejected on this basis 
could reasonably be considered as suf\mbox{}f\mbox{}iciently out-of-line to warrant this treatment. The value of $0.1$ is larger than anyone would 
reasonably choose; however, it was interesting to investigate how our results could change in this rather extreme case.

Table \ref{tab:3-table} shows the values of the seven parameters for the f\mbox{}its to databases selected using these four values of $\mathrm{p}_{min}$. 
The errors shown correspond to $\mathrm{p}_{min}=0.0027$. In fact, when the errors are calculated with the factor $\sqrt{\chi^2/\mathrm{NDF}}$ 
included, they do not vary much with the value of $\mathrm{p}_{min}$. As $\mathrm{p}_{min}$ increases, the database being f\mbox{}itted shrinks and so 
the raw errors increase. However, the factor $\sqrt{\chi^2/\mathrm{NDF}}$ decreases as the f\mbox{}it quality improves (despite the decrease in 
${\mathrm{NDF}}$) and the two ef\mbox{}fects largely compensate. Table \ref{tab:3-table} shows clearly the remarkable stability of the f\mbox{}it as the 
criterion for the rejection of data points is varied. In the case of $G_\rho$, the variation is approximately equal to the uncertainty in its 
determination. In the other six cases, the variation is much smaller than the error. The quality of the f\mbox{}it, judged by standard statistical 
criteria, improves considerably as $\mathrm{p}_{min}$ is increased, but at the cost of the loss of many points from the database. For 
$\mathrm{p}_{min}=0.0027$, there are $668$ entries, $\chi^2=864.8$ and the p-value associated with this $\chi^2$ value is $1.46 \cdot10^{-7}$. At the 
other extreme, for $\mathrm{p}_{min}=0.1$, the database has shrunk to $562$ entries, the value of $\chi^2$ is $591.4$ and the associated p-value is 
$0.138$.

Faced with these results, one can conclude that the combined truncated elastic-scattering database, once the exceptionally bad points have been 
removed, is self-consistent and very robust when additional pruning is done. The output of the analysis is remarkably stable, which suggests that 
nearly all the measurements in the combined truncated elastic-scattering database of $668$ entries are reliable. The apparently poor quality of the 
f\mbox{}it does not seem to be the result of the presence of a substantial number of unreliable points, but rather of a general underestimation of the 
experimental uncertainties, both statistical and normalisation, particularly in the case of the DCS measurements. Any attempt to alter the quoted 
errors would be arbitrary, so we must make judgments on the database as it stands. Our judgment is that it is best to reject as few experimental 
points as possible, by using the value $\mathrm{p}_{min}=0.0027$. We then have to live with a rather poor f\mbox{}it, which we have taken into account by 
increasing the uncertainties by the factor $\sqrt{\chi^2/\mathrm{NDF}}=1.1438$. All the results henceforth correspond to this f\mbox{}it. Any uneasiness 
about this small value of $\mathrm{p}_{min}$ should be put to rest by the observation that, for $\mathrm{p}_{min}=0.01$, the database is very little 
reduced (a few $\pi^+ p$ points are removed) and the output is practically unchanged.

The correlation matrix for the seven parameters of the $\pi N$ model is given in Table \ref{tab:4-table}; the numbers correspond to the f\mbox{}it with
$\mathrm{p}_{min}=0.0027$. This matrix, together with the errors given in Table \ref{tab:3-table}, enables the correct uncertainties to be assigned 
to any quantities calculated from the output of the f\mbox{}it. Table \ref{tab:3-table} shows that the value of $K_\sigma$ is consistent with $0$; the 
quality of the f\mbox{}it would deteriorate very little if, in fact, this parameter were set to $0$. The value of $G_\sigma$ is very little correlated 
with the values of the other f\mbox{}ive parameters. However, those parameters ($G_\rho$, $K_\rho$, $g_{\pi NN}$, $g_{\pi N \Delta}$ and $Z$) are all 
strongly correlated with each other, showing that the contributions of the three processes involved are intimately connected. (As expected, the 
correlations among the model parameters are signif\mbox{}icantly smaller when the f\mbox{}loating of the data sets is not allowed in the f\mbox{}it.)

We see from Table \ref{tab:3-table} that the value of $g_{\pi NN}$ is particularly stable; converted to the usual pseudovector coupling 
constant\footnote{Some authors redef\mbox{}ine $f_{\pi NN}^2$, absorbing in it the factor $4 \pi$.}, our result is
\[
 \frac{f_{\pi NN}^2}{4 \pi}= \left( \frac{\mu_c}{2 m_p} \right)^2 \frac{g_{\pi NN}^2}{4 \pi} =0.0733(14) \, .
\]
As discussed in Section 5.2, the error given may be an underestimate. Until the 1990s, it was generally accepted that the value of $f_{\pi NN}^2/{4\pi}$ 
was around $0.080$, and there have been more recent claims for such high value; for example, the value $0.0808(20)$ was obtained in Ref.~\cite{teoe} (the 
error corresponds to the linear combination of the two uncertainties given there). However, many signif\mbox{}icantly lower values have also appeared in 
the literature. The Nijmegen $NN$ potential uses the value $0.075$; a f\mbox{}it to the deuteron photodisintegration data \cite{jawo} conf\mbox{}irms this 
result. Ref.~\cite{rget} gives the value $0.0732(11)$, while Ref.~\cite{bblm} reports a value as low as $0.071(2)$.

Since the $I=J=0$ t-channel exchange is an ef\mbox{}fective interaction, representing a number of diagrams with loops and the exchange of known scalar 
mesons, the values of $G_\sigma$ and $K_\sigma$ are unique to this analysis. The value of $K_\rho$ from our f\mbox{}it is problematic. It is quite small, 
and considerably larger values come from the study of nucleon form factors, $N N$ scattering and deuteron photodisintegration. This dif\mbox{}ference may 
be a ref\mbox{}lection of the strong correlations, already noted, between the contributions of the $N$-, $\Delta$- and $\rho$-exchange processes, or it 
may be due to the omission of contributions whose ef\mbox{}fect is mainly compensated by a shift in the value of $K_\rho$. In either case, it is 
clear that one must adopt a cautious attitude to the values of ef\mbox{}fective parameters obtained using hadronic models. (The same applies to boson 
exchange models of $N N$ scattering.)

For the $\Delta$ contribution, the values of $g_{\pi N \Delta}$ and $Z$ are very stable. Our way of incorporating the spin-$1/2$ contribution has been 
used for a long time, and our value $Z = -0.53(6)$ is consistent with the much older value $-0.45(20)$ in Ref.~\cite{oo}. The spin-1/2 contribution can 
alternatively be absorbed into contact interactions (see, for example, Ref.~\cite{bhm}), but there would be no advantage in adopting this possibility.

\section{Results for the threshold quantities and hadronic phase shifts}

From the model parameters and their uncertainties given in Table \ref{tab:3-table} for $\mathrm{p}_{min}=0.0027$, as well as the correlation matrix given 
in Table \ref{tab:4-table}, we can calculate the isoscalar and isovector $s$-wave scattering lengths and the isoscalar(isovector)-scalar(vector) $p$-wave 
scattering volumes. The results are
\[
\frac{1}{3}\:\tilde{a}_{0+}^{1/2}+\frac{2}{3}\:\tilde{a}_{0+}^{3/2}=0.0022(12)\: \mu_c^{-1} ,
\]
\[
-\frac{1}{3}\:\tilde{a}_{0+}^{1/2}+\frac{1}{3}\:\tilde{a}_{0+}^{3/2}=-0.07742(61)\: \mu_c^{-1} , \]
\begin{equation}
\frac{1}{3}\:\tilde{a}_{1-}^{1/2}+\frac{2}{3}\:\tilde{a}_{1-}^{3/2}+\frac{2}{3}\:\tilde{a}_{1+}^{1/2}+\frac{4}{3}\:\tilde{a}_{1+}^{3/2}=0.2044(19)\: \mu_c^{-3} ,
\label{eq:atildas}
\end{equation}
\[
-\frac{1}{3}\:\tilde{a}_{1-}^{1/2}+\frac{1}{3}\:\tilde{a}_{1-}^{3/2}-\frac{2}{3}\:\tilde{a}_{1+}^{1/2}+\frac{2}{3}\:\tilde{a}_{1+}^{3/2}=0.1738(18)\: \mu_c^{-3} ,
\]
\[
\frac{1}{3}\:\tilde{a}_{1-}^{1/2}+\frac{2}{3}\:\tilde{a}_{1-}^{3/2}-\frac{1}{3}\:\tilde{a}_{1+}^{1/2}-\frac{2}{3}\:\tilde{a}_{1+}^{3/2}=-0.1839(19)\: \mu_c^{-3} ,
\]
\[
-\frac{1}{3}\:\tilde{a}_{1-}^{1/2}+\frac{1}{3}\:\tilde{a}_{1-}^{3/2}+\frac{1}{3}\:\tilde{a}_{1+}^{1/2}-\frac{1}{3}\:\tilde{a}_{1+}^{3/2}=-0.06780(83)\: \mu_c^{-3} .
\]

Converting these results to the familiar spin-isospin quantities, we obtain
\[
\tilde{a}_{0+}^{3/2}=-0.0752(16)\: \mu_c^{-1} ,\qquad \tilde{a}_{0+}^{1/2}=0.1571(13)\: \mu_c^{-1} ,
\]
\begin{equation}
\tilde{a}_{1-}^{3/2}=-0.04176(80)\: \mu_c^{-3} ,\qquad \tilde{a}_{1-}^{1/2}=-0.0799(16)\: \mu_c^{-3} ,
\label{eq:atildasvalues}
\end{equation}
\[
\tilde{a}_{1+}^{3/2}=0.2100(20)\: \mu_c^{-3} ,\qquad \tilde{a}_{1+}^{1/2}=-0.03159(67)\: \mu_c^{-3} .
\]

Our results for the $s$-wave scattering lengths $\tilde{a}_{0+}^{3/2}$ and $\tilde{a}_{0+}^{1/2}$ in Eqs.~(\ref{eq:atildasvalues}) may seem surprising, 
yet they are almost identical to those obtained in Refs.~\cite{fm,m}; these values have been very stable for over ten years. The large quantity of data 
below $100$ MeV obtained at pion factories since 1980, when analysed separately, leads to results for the $s$-wave scattering lengths (and hadronic phase 
shifts) which are signif\mbox{}icantly dif\mbox{}ferent from those extracted by using dispersion relations and databases extending up to the GeV region. 
The dif\mbox{}ferences amount to about $10 \%$, with uncertainties which are much smaller, around $2 \%$ or less. There is another dif\mbox{}ference 
which is more serious still. From the results in Eqs.~(\ref{eq:atildasvalues}), one obtains
\[
 \tilde{a}^{cc} = \frac{2}{3} \, \tilde{a}_{0+}^{1/2} + \frac{1}{3} \, \tilde{a}_{0+}^{3/2} = 0.0797(11)\: \mu_c^{-1} \, .
\]
On the other hand, the value of $\tilde{a}^{cc}$ derived from the experiment of Ref.~\cite{ss} on pionic hydrogen, using an EM correction obtained from 
a potential model, is $0.0883(8)$ $\mu_c^{-1}$. The EM correction used in deriving this result employed simple potential forms, and made only a rough 
estimate of the ef\mbox{}fect of the $\gamma n$ channel. A full three-channel calculation \cite{orwmg}, using potentials of the same form as those used 
for the corrections of Refs.~\cite{gmorw,gmorww}, results in a change in the EM correction used in Ref.~\cite{ss}, which reduces the value of 
$\tilde{a}^{cc}$. However, even with this change, there is still a very large dif\mbox{}ference between the result from the PSA and that from pionic 
hydrogen, which cannot be explained by appealing to the violation of isospin invariance in the em-modif\mbox{}ied hadronic interaction.

It is dif\mbox{}f\mbox{}icult to account for this discrepancy. If the single value of $\tilde{a}^{cc}$ from pionic hydrogen were to be added to the 
combined database of $668$ points, it would be an immediate candidate for rejection. It is therefore necessary to look at possible problems with the 
combined elastic-scattering database. We considered whether to follow other $\pi N$ analyses, which took into account only the shapes of the FRANK83 
data sets and ignored their absolute normalisations. There was vigorous debate for many years about the reliability of the normalisations of these 
eight data sets. However, the experimental group has neither withdrawn its DCS results, nor hinted that its normalisations might be in error. Moreover, 
comparison with the normalisation uncertainties quoted for the other DCS experiments listed in Tables \ref{tab:1-table} and \ref{tab:2-table} shows that 
the experimental group quoted quite generous uncertainties. Inspection of the scale factors $z_j$ in Tables \ref{tab:1-table} and \ref{tab:2-table} for 
the eight data sets shows that a decision to freely f\mbox{}loat the FRANK83 data sets is neither called for nor suggested in our analysis. We therefore 
made the decision to accept these data sets at their face value, as we did for all the other elastic-scattering measurements. In fact, when our analysis 
is repeated, with the only change being the free f\mbox{}loating of the FRANK83 data sets, the value of $\tilde{a}^{cc}$ changes by only a very small 
amount, to $0.0794(11)$ $\mu_c^{-1}$.

We have tested the ef\mbox{}fect of one modif\mbox{}ication to the database. Assuming that the normalisation uncertainties of most of the experiments may 
have been underestimated, we doubled them all. However, the f\mbox{}it then became unstable and it was impossible to obtain reliable results. There seems 
to be a delicate balance between the statistical and the normalisation errors which makes a stable f\mbox{}it possible. We hope in time to explore more 
sophisticated ways of selectively modifying the database, but the def\mbox{}initive resolution of the present discrepancy may require a reappraisal by 
experimenters of the whole body of low-energy elastic-scattering data.

Because of the violation of isospin invariance in the em-modif\mbox{}ied hadronic interaction (see next section), it is important to give also the results 
for the $s$-wave scattering length and $p$-wave scattering volumes obtained from the analysis of the truncated $\pi^+ p$ database alone (Table 
\ref{tab:1-table}), using the parametric forms in Eqs.~(\ref{eq:delta0+3/2})-(\ref{eq:delta1+3/2}). For these quantities we shall not use the 
superscript $3/2$, since isospin invariance is no longer assumed to hold; we use instead the superscript `$\pi^+ p$'. With this notation, we obtain
\begin{equation} \label{eq:atilda+values}
\tilde{a}_{0+}^{\pi^+ p}=-0.0751(39)\: \mu_c^{-1},\;
\tilde{a}_{1-}^{\pi^+ p}=-0.0526(53)\: \mu_c^{-3},\;
\tilde{a}_{1+}^{\pi^+ p}=0.2013(35)\: \mu_c^{-3}.
\end{equation}
The results in Eqs.~(\ref{eq:atilda+values}) agree well with those of Ref.~\cite{fm}.

\begin{figure}
\begin{center}
\includegraphics [width=15.5cm] {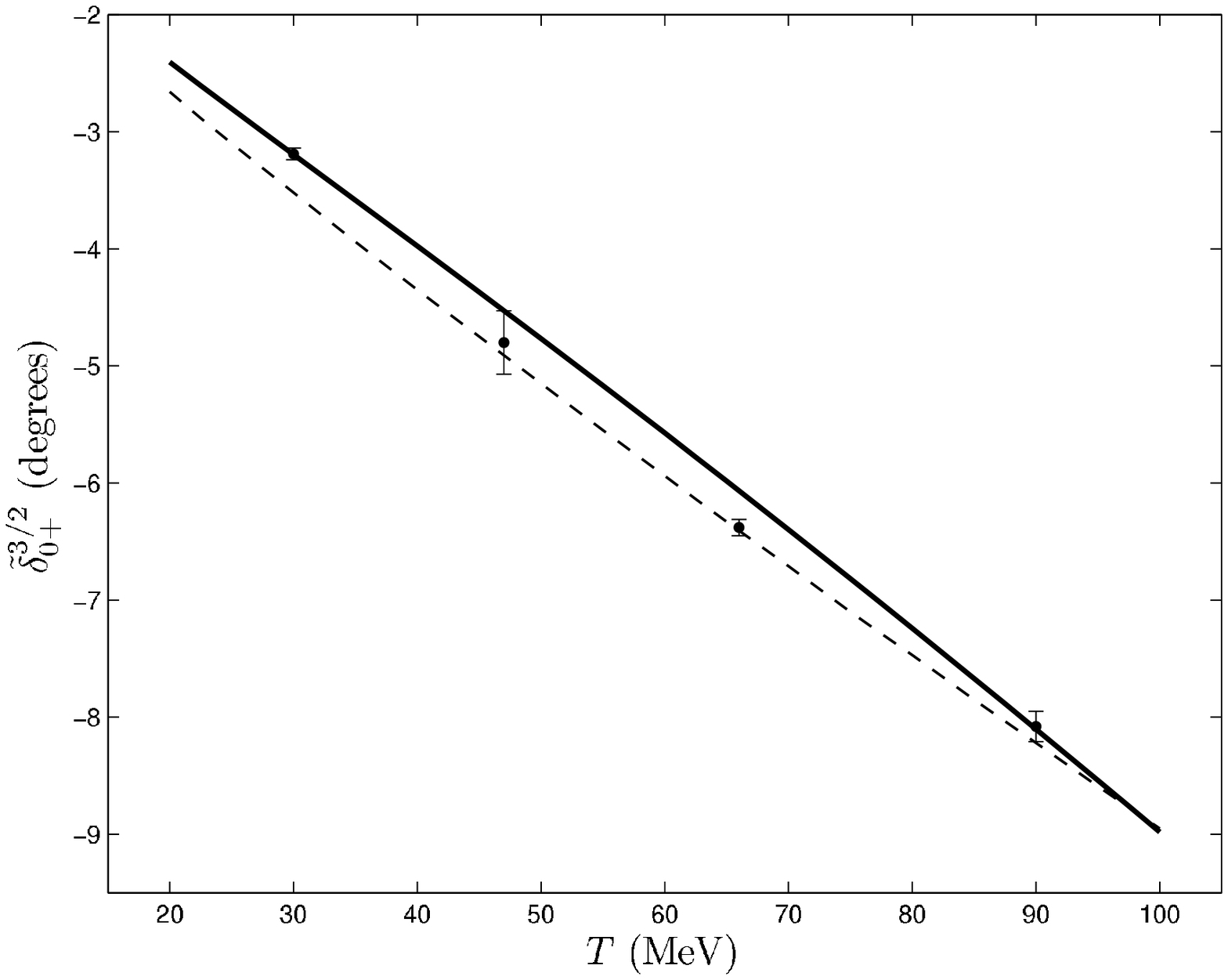}
\caption{\label{fig:a}The em-modif\mbox{}ied hadronic phase shift $\tilde\delta_{0+}^{3/2}$} from the present work (solid curve) and from the current 
GWU solution \cite{abws} (dashed curve). The four single-energy points of Ref.~\cite{abws}, at $30, 47, 66$ and $90$ MeV, are also shown.
\end{center}
\end{figure}

\begin{figure}
\begin{center}
\includegraphics [width=15.5cm] {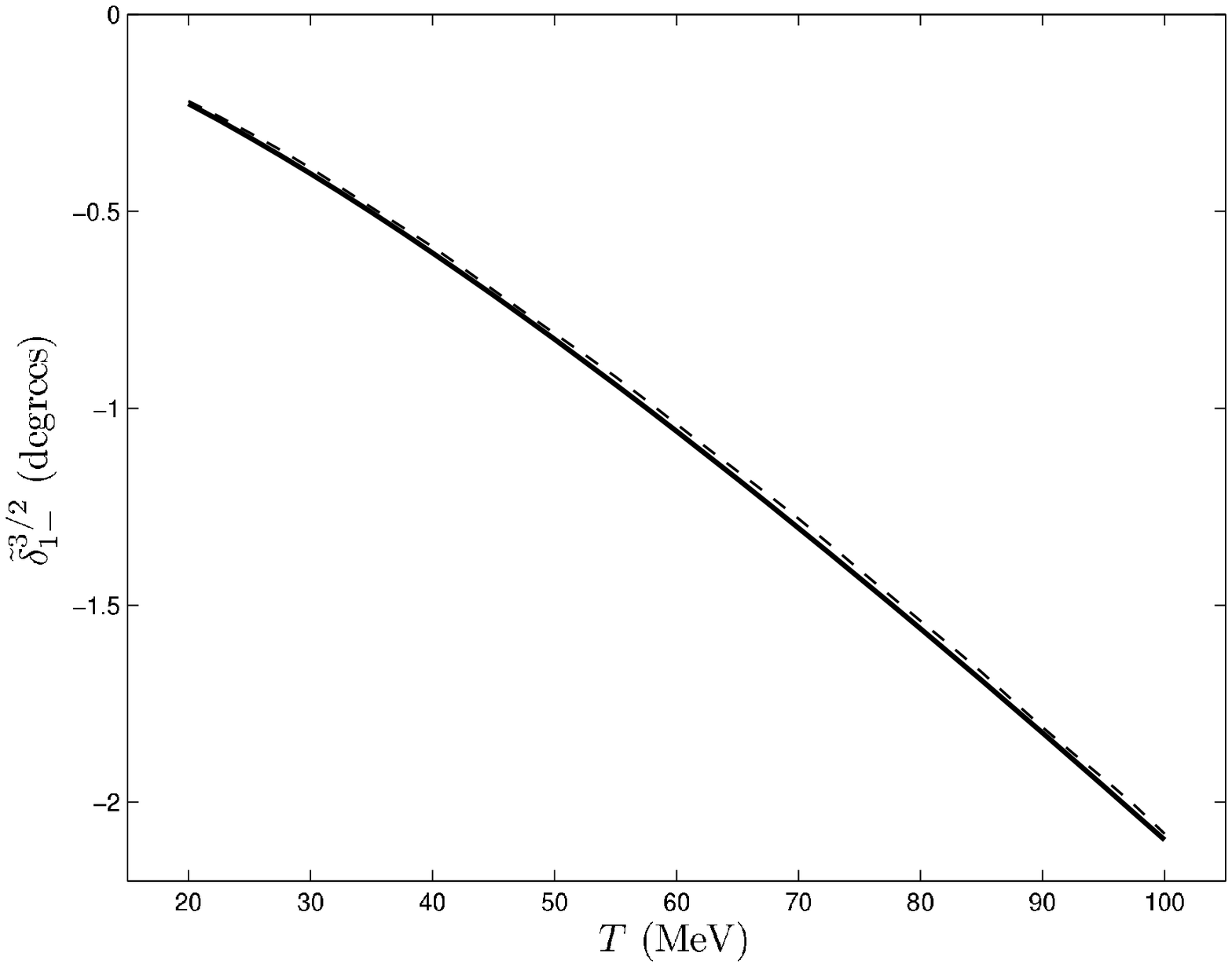}
\caption{\label{fig:b}The em-modif\mbox{}ied hadronic phase shift $\tilde\delta_{1-}^{3/2}$} from the present work (solid curve) and from the current 
GWU solution \cite{abws}
(dashed curve).
\end{center}
\end{figure}

\begin{figure}
\begin{center}
\includegraphics [width=15.5cm] {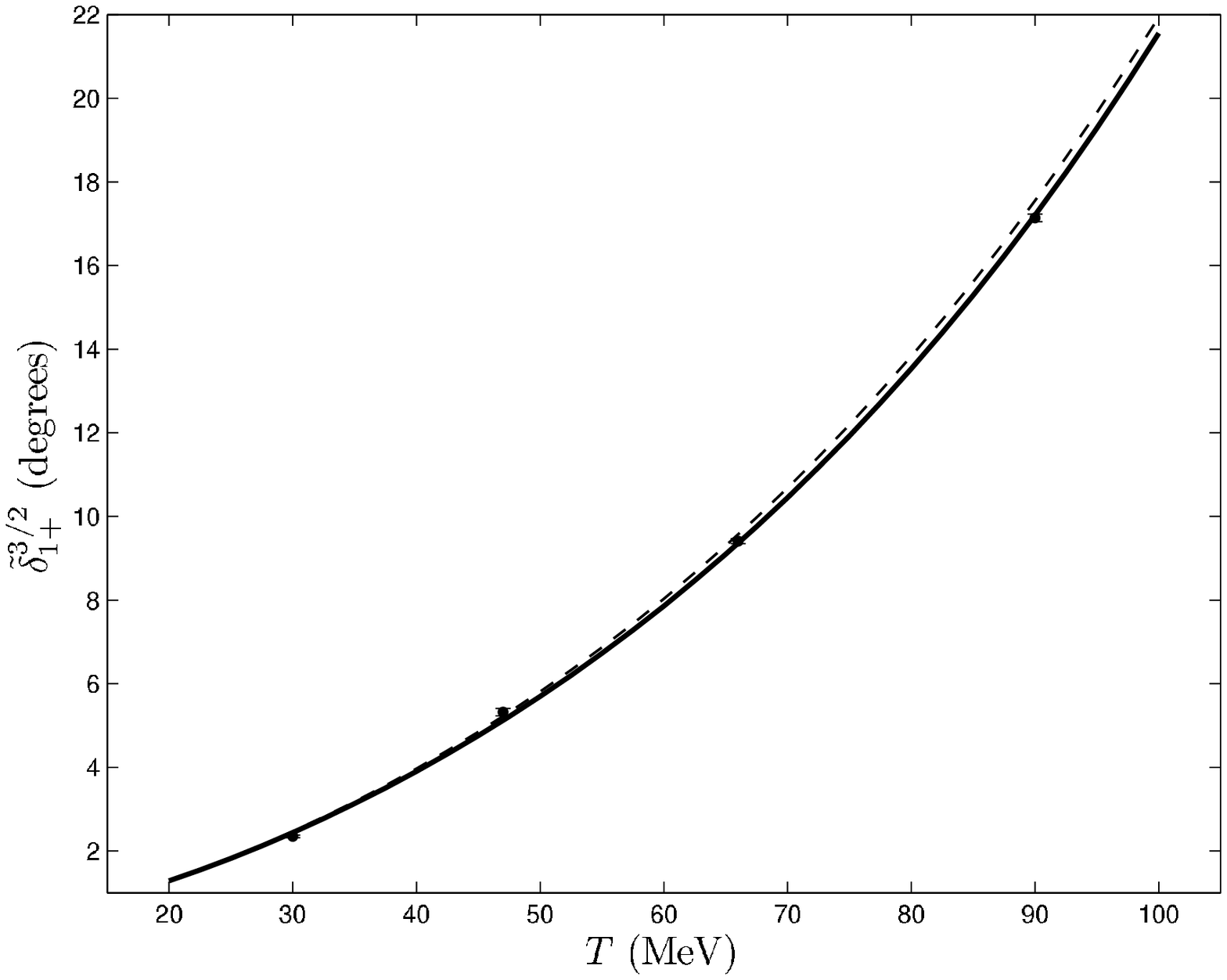}
\caption{\label{fig:c}The em-modif\mbox{}ied hadronic phase shift $\tilde\delta_{1+}^{3/2}$} from the present work (solid curve) and from the current 
GWU solution \cite{abws} (dashed curve). The four single-energy points of Ref.~\cite{abws}, at $30, 47, 66$ and $90$ MeV, are also shown.
\end{center}
\end{figure}

\begin{figure}
\begin{center}
\includegraphics [width=15.5cm] {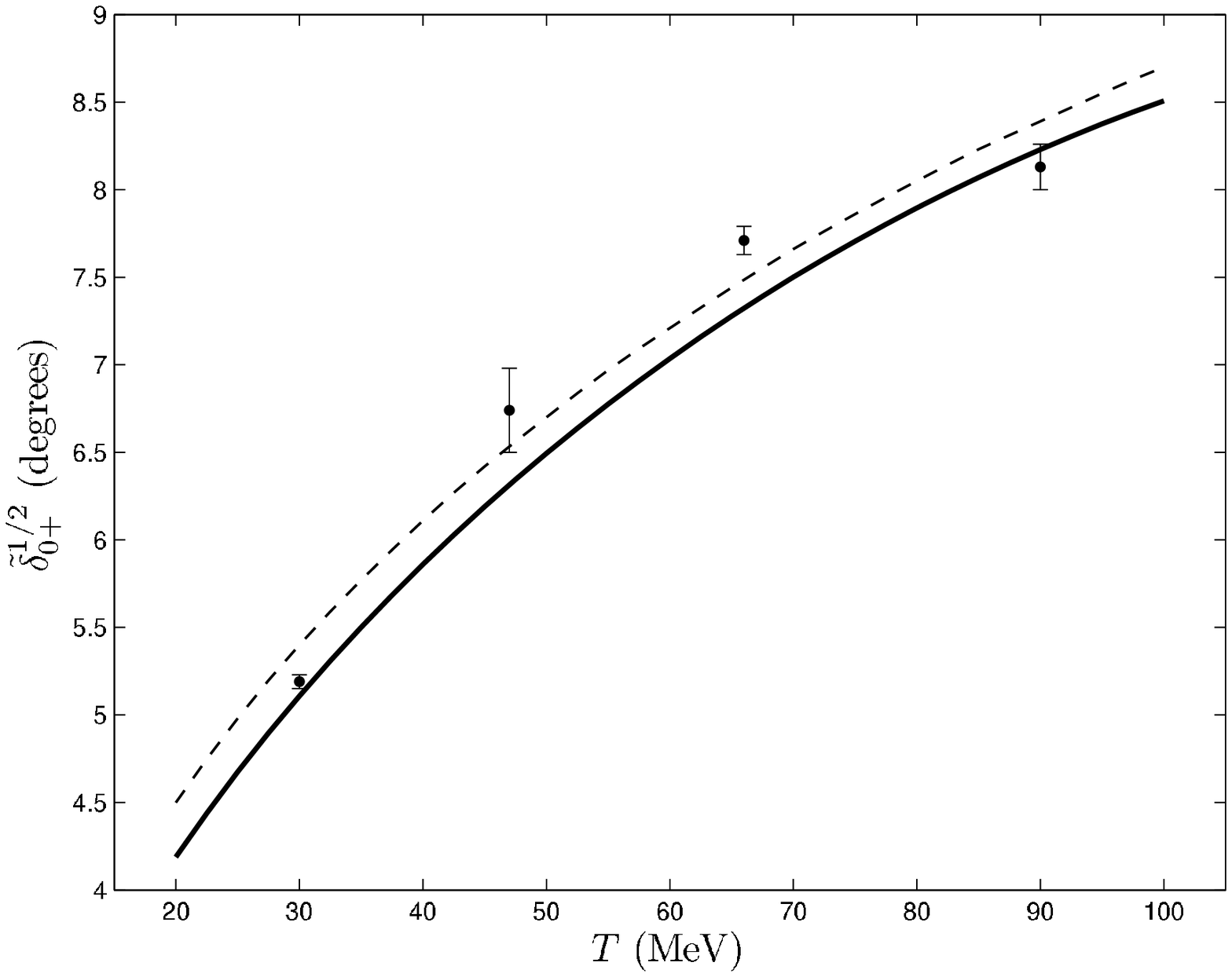}
\caption{\label{fig:d}The em-modif\mbox{}ied hadronic phase shift $\tilde\delta_{0+}^{1/2}$} from the present work (solid curve) and from the current 
GWU solution \cite{abws} (dashed curve). The four single-energy points of Ref.~\cite{abws}, at $30, 47, 66$ and $90$ MeV, are also shown.
\end{center}
\end{figure}

\begin{figure}
\begin{center}
\includegraphics [width=15.5cm] {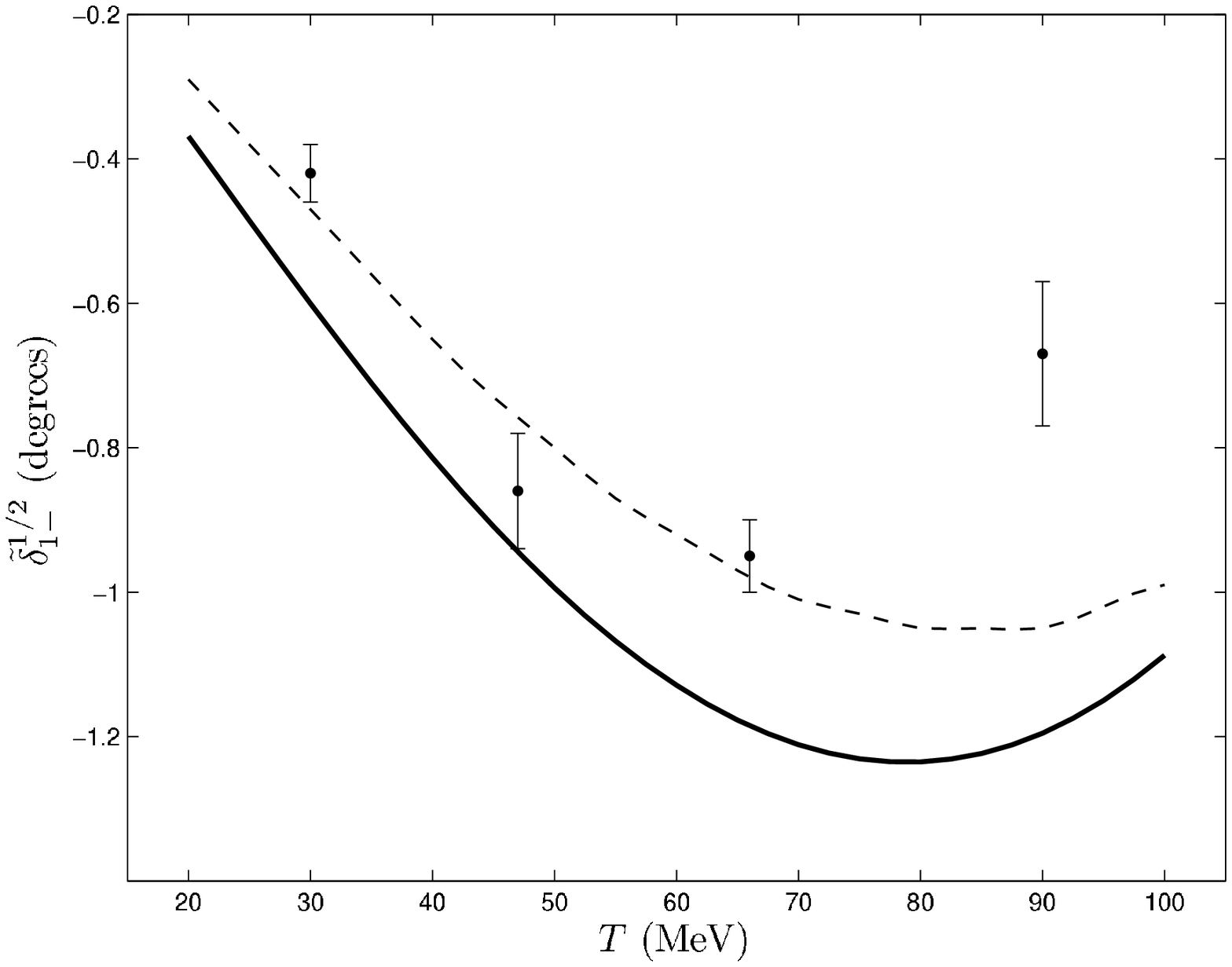}
\caption{\label{fig:e}The em-modif\mbox{}ied hadronic phase shift $\tilde\delta_{1-}^{1/2}$} from the present work (solid curve) and from the current 
GWU solution \cite{abws} (dashed curve). The four single-energy points of Ref.~\cite{abws}, at $30, 47, 66$ and $90$ MeV, are also shown.
\end{center}
\end{figure}

\begin{figure}
\begin{center}
\includegraphics [width=15.5cm] {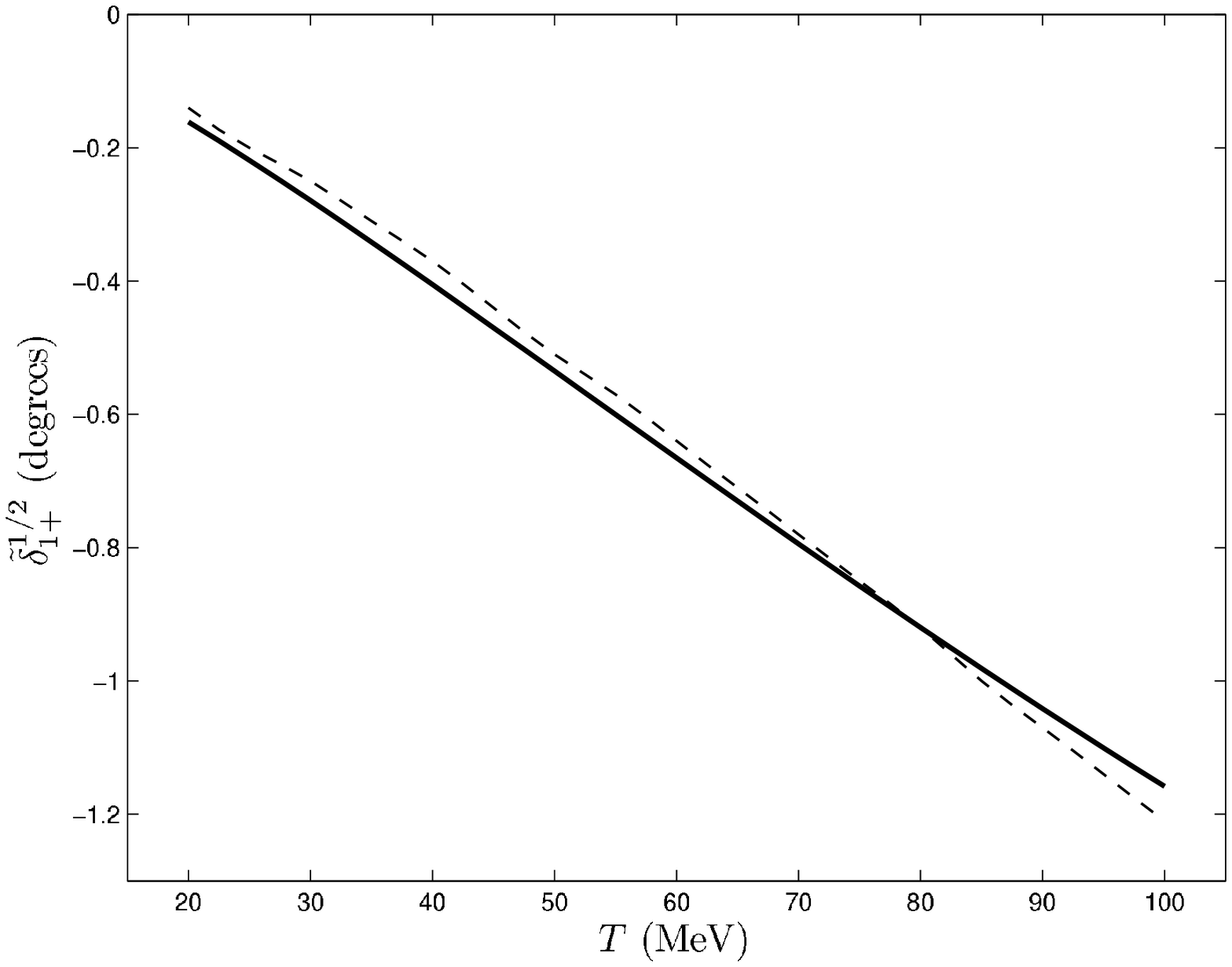}
\caption{\label{fig:f}The em-modif\mbox{}ied hadronic phase shift $\tilde\delta_{1+}^{1/2}$} from the present work (solid curve) and from the current 
GWU solution \cite{abws} (dashed curve).
\end{center}
\end{figure}

The f\mbox{}inal $s$- and $p$-wave em-modif\mbox{}ied hadronic phase shifts, from the model f\mbox{}it to the combined truncated elastic-scattering 
database, are given in Table \ref{tab:5-table}, in $5$ MeV intervals from $20$ to $100$ MeV. These hadronic phase shifts are also shown in 
Figs.~\ref{fig:a}-\ref{fig:f}, together with the current GWU solution \cite{abws} and their four single-energy values (whenever available).

It is evident from Figs.~\ref{fig:a} and \ref{fig:d} that our values of the $s$-wave hadronic phase shifts $\tilde\delta_{0+}^{3/2}$ and 
$\tilde\delta_{0+}^{1/2}$ dif\mbox{}fer signif\mbox{}icantly from the GWU results. Our values of $\tilde\delta_{0+}^{3/2}$ are less negative, but 
converge towards the GWU values as the energy approaches $100$ MeV. Interestingly, the GWU single-energy results at $30$ and $90$ MeV agree with our 
results. For $\tilde\delta_{0+}^{1/2}$, our values are consistently smaller, with a slight convergence towards $100$ MeV. Once again, the GWU 
single-energy results at $30$ and $90$ MeV agree with ours, but the ones at $47$ and $66$ MeV do not. For the $p$-wave hadronic phase shifts 
$\tilde\delta_{1-}^{3/2}$, $\tilde\delta_{1+}^{3/2}$ and $\tilde\delta_{1+}^{1/2}$, inspection of Figs.~\ref{fig:b}, \ref{fig:c} and \ref{fig:f} 
shows that there is good agreement between the two solutions. The signif\mbox{}icant dif\mbox{}ference in the $p$-wave part of the interaction occurs 
for $\tilde\delta_{1-}^{1/2}$. Our values are clearly considerably lower at all energies and the single-energy GWU results are a long way from ours 
except at $47$ MeV. The dif\mbox{}ferences between our results for the hadronic phase shifts and those of GWU are not due to the improved stage 1 EM 
corrections which we have used. In fact, the exact values of the EM corrections used have very little ef\mbox{}fect on the output of the PSA. The 
dif\mbox{}ferences are due almost entirely to the method we have used, restricting the data being analysed to $T \leq 100$ MeV and freely f\mbox{}loating 
the data from only three experiments, thus respecting as far as possible the measurements as they have been published.

We conclude this section with a discussion of the measurements of Ref.~\cite{meier} which have not been included in our database. Using our phase-shift 
output, we have created Monte-Carlo predictions for the analysing power corresponding to each of their $28$ experimental points. For the three 
experimental data sets, the resulting values of $\chi^2_{min}$ are $12.72$, $7.41$ and $15.87$, for $12$, $6$ and $10$ degrees of freedom. The values 
of the scale factor for the three sets (in the same order) are $1.0079$, $0.9718$ and $1.0360$. It is evident, not only that all the data points satisfy 
the acceptance criterion which was applied to the full $\pi^+ p$ and $\pi^- p$ databases (see Section 4), but also that our hadronic phase shifts 
reproduce the data of Ref.~\cite{meier} very well; even the data set which is reproduced least well corresponds to a p-value of $0.1035$. Our conclusion 
is that, even if the data of Ref.~\cite{meier} were in a form which could easily enable their inclusion in our database, the impact on our results would 
have been negligible. For one thing, the data are very well reproduced by our present solution; for another, they correspond to about $4.2 \%$ of the 
combined truncated elastic-scattering database which we have used for our analysis.

\section{Analysis of the $\pi^- p$ charge-exchange data and the violation of isospin invariance}

For reasons given in Section 1, our PSA was based on the elastic-scattering data only. The EM corrections of Refs.~\cite{gmorw,gmorww} were applied and 
the hadronic part of each scattering amplitude was parameterised in a framework of formal isospin invariance, leading to the extraction of the six $s$- 
and $p$-wave em-modif\mbox{}ied hadronic phase shifts $\tilde\delta$. Because of the residual EM ef\mbox{}fects, it is not very surprising that there is 
evidence for the violation of isospin invariance at the em-modif\mbox{}ied level \cite{glk,m}. In Ref.~\cite{glk}, it was shown that the violation of 
isospin invariance appears mainly in the $s$-wave amplitude, with some ef\mbox{}fect also present in the no-spin-f\mbox{}lip $p$ wave. We will now 
strengthen the evidence in Refs.~\cite{glk,m}, by using the output of our PSA of the elastic-scattering data to attempt to reproduce the measurements 
on the CX reaction $\pi^- p \to \pi^0 n$.

There is an extensive database of CX measurements below $100$ MeV. The modern measurements of the DCS come from four experiments, namely FITZGERALD86 
\cite{dhf}, FRLE{\v Z}98 \cite{efr}, ISENHOWER99 \cite{ldi} and SADLER04 \cite{mesa}. The FITZGERALD86 data comprise measurements of the DCS close to 
$0^\circ$ at seven energies, from $32.5$ to $63.2$ MeV; only their direct measurements were used here (their extrapolated values to $0^\circ$ were not 
taken into account). FRLE{\v Z}98 measured the DCS at $27.5$ MeV, at six angles between $4.7^\circ$ and $50.9^\circ$. ISENHOWER99 measured the DCS at 
$10.6$, $20.6$ and $39.4$ MeV; the groups of points near $0^\circ$, $90^\circ$ and $180^\circ$ have independent beam normalisations, thus leading to 
eight independent data sets. SADLER04 made detailed measurements of the DCS at $63.9$, $83.5$ and $94.6$ MeV. These four experiments share a common 
characteristic: they all measured the CX DCS in the forward hemisphere. We will now explain why they are expected to be more conclusive than the rest 
of the experiments on the issue of isospin invariance. The main contributions to the CX scattering amplitude in the low-energy region come from the 
real parts of the $s$ and $p$ waves. Taking into account the opposite signs of these contributions, the two main components of the scattering amplitude 
almost cancel each other in the forward direction around $40$ MeV, thus enabling small ef\mbox{}fects to show up. The destructive-interference phenomenon 
acts like a magnifying glass, probing the smaller components in the $\pi N$ dynamics. Note that the two studies \cite{glk,m} of the isospin-invariance 
violation did not have the results in Refs.~\cite{efr}-\cite{mesa} available to them. 

There are a number of additional CX data sets which, for various reasons, are not expected to contribute to the discussion of the isospin-invariance 
violation. a) The experiment of DUCLOS73 \cite{duc} measured the DCS near $180^\circ$ at $22.6$, $32.9$ and $42.6$ MeV. Apart from the large statistical 
uncertainties of these three data points (which, according to Eq.~(\ref{eq:zj}), bring the scale factors closer to $1$), the contributions of the $s$ and 
$p$ waves {\it add} in the backward hemisphere (constructive interference), thus masking any small ef\mbox{}fects which might be present in the $\pi N$ 
dynamics at these energies. b) There are two similar experiments, SALOMON84 \cite{smpr} and BAGHERI88 \cite{bagh}, whose output consisted of the 
f\mbox{}irst three coef\mbox{}f\mbox{}icients in a Legendre expansion of the DCS at six energies between $27.4$ and $91.7$ MeV. Even if the correlation 
matrices, corresponding to the dif\mbox{}ferent energies where the measurements were taken, could be found in these papers, the raw DCS data could not be 
reconstructed; one could obtain only the f\mbox{}itted values, but no knowledge of how these f\mbox{}itted values compare with the measured values of the 
DCS. c) There is a measurement of the total cross section for the CX reaction at $90.9$ MeV from the experiment of BUGG71 \cite{bbdscw}. d) Finally, there 
are measurements of the analysing power at $100$ MeV (STA{\v S}KO93 \cite{stasko}) and $98.1$ MeV (GAULARD99 \cite{gaul}). In cases (c) and (d) above, the 
energy used was high; additionally, in case (d), the sensitivity of the analysing power to the ef\mbox{}fect being investigated in this section is 
questionable.

The full database for the CX reaction consists of $31$ data sets, containing $159$ data points. In cases where a normalisation uncertainty was not 
properly reported, we had to assign realistic normalisation errors by comparison with those quoted for other experiments. The precise details are not 
important.

The f\mbox{}irst step was to check the consistency of the CX data in the same way as we have done for the $\pi^- p$ elastic-scattering database. The 
$I=3/2$ amplitudes were f\mbox{}ixed from the f\mbox{}inal f\mbox{}it to the $\pi^+ p$ database of Table \ref{tab:1-table} and the seven parameters for 
the $I=1/2$ amplitudes were varied in order to optimise the description of the CX data via the minimisation of the Arndt-Roper function. The minimum value 
of $\chi^2$ was $154.9$ for $152$ degrees of freedom (with p-value equal to $0.42$); no doubtful data sets or individual data points could be found. On 
the other hand, when we attempted to reproduce the CX database using the output of the f\mbox{}it to the combined truncated elastic-scattering data, the 
value of $\chi^2$ jumped to an enormous $508.9$. The DCS of Refs.~\cite{dhf}-\cite{duc} and the total cross section of Ref.~\cite{bbdscw}, $131$ data 
points in total, contribute $471.2$ units to the overall $\chi^2$; as previously mentioned, it is in the DCS measurements in the forward hemisphere that 
the ef\mbox{}fect is expected to show up best. The analysing-power measurements of Refs.~\cite{stasko,gaul} are reproduced well. With the exception of the 
data set at $45.6$ MeV, the indirect data of Refs.~\cite{smpr,bagh} are also well reproduced.

The conclusion is that the CX database cannot be satisfactorily reproduced from the PSA of the $\pi^\pm p$ elastic-scattering data which used a framework 
of formal isospin invariance. This is conclusive evidence that isospin invariance is violated at the em-modif\mbox{}ied hadronic level, thus corroborating 
the f\mbox{}indings of Refs.~\cite{glk,m} which were based on a signif\mbox{}icantly smaller CX database. The predictions, based on the output from the 
PSA of the elastic-scattering data, seriously underestimate the measured CX cross sections.

We content ourselves here with this clear evidence for the violation of isospin invariance at the em-modif\mbox{}ied level. What is still needed is a new 
PSA, of the combined $\pi^+ p$ elastic-scattering and $\pi^- p$ CX databases, using the same model as before. The output will consist of $I = 3/2$ 
hadronic phase shifts, which one expects to be very close to those given in Section 6, and new `I = 1/2' hadronic phase shifts, which will be 
substantially dif\mbox{}ferent from those given in Section 6. The dif\mbox{}ferences will be able to make more precise the conclusions of 
Ref.~\cite{glk}, on the size of the violation of isospin invariance in the $s$ wave and the two $p$ waves. Further, comparison of the values of the 
$s$-wave threshold parameter for $\pi^-p$ CX scattering, obtained from the new PSA and from the width of the $1s$ level of pionic hydrogen, will provide 
further information on the reliability of the $\pi^+ p$ elastic-scattering and $\pi^- p$ CX databases.

\section{Discussion}

In the present paper, we have reported the results of a PSA of the $\pi^\pm p$ elastic-scattering data for $T\leq 100$ MeV using the recently obtained 
EM corrections of Refs.~\cite{gmorw,gmorww}; the analysis was performed with a hadronic interaction described within a framework of formal isospin 
invariance. We found that it was possible to obtain self-consistent databases by removing the measurements of only two $\pi^+ p$ data sets and one 
$\pi^- p$ set, as well as a very small number of single data points; the removal of these outliers resulted in enormous reductions in the minimum value 
of $\chi^2$ for the separate f\mbox{}its to the two elastic-scattering databases.

The $\pi N$ model of Ref.~\cite{glmbg}, based on s- and u-channel diagrams with $N$ and $\Delta$ in the intermediate states, and $\sigma$ and $\rho$ 
t-channel exchanges, was subsequently f\mbox{}itted to the elastic-scattering database obtained after the removal of the outliers. The model-parameter 
values showed an impressive stability when subjected to dif\mbox{}ferent criteria for the rejection of experiments (see Table \ref{tab:3-table}); we 
f\mbox{}inally adopted the criterion which removes the smallest amount of experimental data, and adjusted the output uncertainties in such a way as to 
take account of the quality of the f\mbox{}it. Our f\mbox{}inal result for the pseudovector $\pi N N$ coupling constant is $0.0733 \pm 0.0014$. Our $s$- 
and $p$-wave em-modif\mbox{}ied hadronic phase shifts are given in Table \ref{tab:5-table}. Big dif\mbox{}ferences in the $s$-wave part of the interaction 
were found when comparing our hadronic phase shifts with the current GWU solution \cite{abws} (see Figs.~\ref{fig:a} and \ref{fig:d}); there is general 
agreement in the $p$ waves, except for the $\tilde\delta_{1-}^{1/2}$ hadronic phase shift. We observed that these dif\mbox{}ferences come from our 
decision to restrict the analysis to data for $T \leq 100$ MeV, and pointed out the apparent mismatch between this data and data at higher energies. We 
also showed that there is a serious discrepancy between our $s$-wave scattering lengths and the value of the $s$-wave threshold parameter for $\pi^- p$ 
elastic scattering obtained from pionic hydrogen. There is no simple way to account for these dif\mbox{}ferences, and serious questions about the 
low-energy elastic-scattering database remain unanswered.

We showed that the experimental results for the CX reaction $\pi^- p \to \pi^0 n$ cannot be reproduced using the output from the PSA of the 
elastic-scattering data; this inability corroborates the f\mbox{}indings of Refs.~\cite{glk,m} concerning the violation of isospin invariance in the 
em-modif\mbox{}ied hadronic $\pi N$ interaction at low energies. We pointed out in Section 1 the need for stage 2 EM corrections (to remove the EM 
ef\mbox{}fects in the hadronic interaction) to be calculated, f\mbox{}irst for all threshold parameters, and subsequently for the analysis of scattering 
data at nonzero energies for all three reaction types. Until such corrections exist, it is necessary to use the existing stage 1 corrections to 
analyse the available scattering data, as we have done.

\begin{ack}
One of us (E.M.) acknowledges a helpful discussion with H.J. Leisi.
\end{ack}

\section*{Note added in proof}

While this work was being reviewed, additional CX data appeared \cite{breit}, namely values of the total cross section at eighteen energies, nine of them 
below $100$ MeV. The measurements below $100$ MeV are reproduced very well by our phase-shift solution.

\newpage
\begin{table}[h!]
{\bf \caption{\label{tab:1-table}}}The data sets comprising the truncated database for $\pi^+ p$ elastic scattering, the pion laboratory kinetic
energy $T$ (in MeV), the number of degrees of freedom $(NDF)_j$ for each set, the scale factor $z_j$ which minimises $\chi_j^2$, the values of
$(\chi_j^2)_{min}$ and the p-value for each set.
\vspace{0.2cm}
\begin{center}
\begin{tabular}{|l|lrlll|l|}
\hline
Data set & $T$ & $(NDF)_j$ & $z_j$ & $\chi_j^2$ & $p$ & Comments \\
\hline
AULD79 & 47.9 & 11 & 1.0101 & 16.4245 & 0.1261 & \\
RITCHIE83 & 65.0 & 8 & 1.0443 & 17.4019 & 0.0262 & \\
RITCHIE83 & 72.5 & 10 & 1.0061 & 4.7745 & 0.9057 & \\
RITCHIE83 & 80.0 & 10 & 1.0297 & 19.3025 & 0.0366 & \\
RITCHIE83 & 95.0 & 10 & 1.0315 & 12.4143 & 0.2583 & \\
FRANK83 & 29.4 & 28 & 0.9982 & 17.4970 & 0.9381 & \\
FRANK83 & 49.5 & 28 & 1.0402 & 34.3544 & 0.1894 & \\
FRANK83 & 69.6 & 27 & 0.9290 & 23.3002 & 0.6688 & \\
FRANK83 & 89.6 & 27 & 0.8618 & 29.1920 & 0.3517 & \\
BRACK86 & 66.8 & 4 & 0.8914 & 2.5020 & 0.6443 & freely f\mbox{}loated \\
BRACK86 & 86.8 & 8 & 0.9387 & 16.6304 & 0.0342 & freely f\mbox{}loated \\
BRACK86 & 91.7 & 5 & 0.9734 & 11.9985 & 0.0348 & \\
BRACK86 & 97.9 & 5 & 0.9714 & 7.5379 & 0.1836 & \\
BRACK88 & 66.8 & 6 & 0.9469 & 11.2438 & 0.0811 & \\
BRACK88 & 66.8 & 6 & 0.9561 & 9.8897 & 0.1294 & \\
WIEDNER89 & 54.3 & 19 & 0.9851 & 14.7822 & 0.7363 & \\
BRACK90 & 30.0 & 6 & 1.0946 & 17.7535 & 0.0069 & \\
BRACK90 & 45.0 & 8 & 1.0055 & 7.8935 & 0.4439 & \\
BRACK95 & 87.1 & 8 & 0.9733 & 13.3922 & 0.0991 & \\
BRACK95 & 98.1 & 8 & 0.9820 & 14.8872 & 0.0614 & \\
JORAM95 & 45.1 & 9 & 0.9548 & 22.2169 & 0.0082 & one point removed \\
JORAM95 & 68.6 & 9 & 1.0503 & 8.8506 & 0.4512 & \\
JORAM95 & 32.2 & 19 & 1.0087 & 23.7410 & 0.2063 & one point removed \\
JORAM95 & 44.6 & 18 & 0.9503 & 29.8018 & 0.0394 & two points removed \\
SEVIOR89 & 98.0 & 6 & 1.0178 & 5.3726 & 0.4970& \\
\hline
\end{tabular}
\end{center}
\end{table}

\newpage
\begin{table*}
{\bf Table 1 continued}
\vspace{0.2cm}
\begin{center}
\begin{tabular}{|l|lrlll|l|}
\hline
Data set & $T$ & $(NDF)_j$ & $z_j$ & $\chi_j^2$ & $p$ & Comments \\
\hline
WIESER96 & 68.34 & 3 & 0.8945 & 2.4732 & 0.4802 & \\
WIESER96 & 68.34 & 4 & 0.9252 & 3.6315 & 0.4582 & \\
KRISS97 & 39.8 & 1 & 1.0121 & 1.7222 & 0.1894 & \\
KRISS97 & 40.5 & 1 & 1.0017 & 0.1400 & 0.7083 & \\
KRISS97 & 44.7 & 1 & 1.0020 & 0.0415 & 0.8385 & \\
KRISS97 & 45.3 & 1 & 1.0025 & 0.0518 & 0.8200 & \\
KRISS97 & 51.1 & 1 & 1.0240 & 3.3209 & 0.0684 & \\
KRISS97 & 51.7 & 1 & 1.0024 & 0.0397 & 0.8421 & \\
KRISS97 & 54.8 & 1 & 1.0068 & 0.1376 & 0.7107 & \\
KRISS97 & 59.3 & 1 & 1.0252 & 1.2497 & 0.2636 & \\
KRISS97 & 66.3 & 2 & 1.0501 & 4.0858 & 0.1297 & \\
KRISS97 & 66.8 & 2 & 1.0075 & 0.5897 & 0.7446 & \\
KRISS97 & 80.0 & 1 & 1.0142 & 0.3704 & 0.5428 & \\
KRISS97 & 89.3 & 1 & 1.0079 & 0.2849 & 0.5935 & \\
KRISS97 & 99.2 & 1 & 1.0550 & 4.1084 & 0.0427 & \\
FRIEDMAN99 & 45.0 & 1 & 1.0423 & 2.1509 & 0.1425 & \\
FRIEDMAN99 & 52.1 & 1 & 1.0172 & 0.2461 & 0.6198 & \\
FRIEDMAN99 & 63.1 & 1 & 1.0364 & 0.4918 & 0.4831 & \\
FRIEDMAN99 & 67.45 & 2 & 1.0524 & 1.2636 & 0.5316 & \\
FRIEDMAN99 & 71.5 & 2 & 1.0501 & 0.8458 & 0.6551 & \\
FRIEDMAN99 & 92.5 & 2 & 1.0429 & 0.5860 & 0.7460 & \\
CARTER71 & 71.6 & 1 & 1.0933 & 2.7422 & 0.0977 & \\
CARTER71 & 97.4 & 1 & 1.0495 & 0.6856 & 0.4077 & \\
PEDRONI78 & 72.5 & 1 & 1.0125 & 0.1416 & 0.7067 & \\
PEDRONI78 & 84.8 & 1 & 1.0319 & 0.3443 & 0.5574 & \\
PEDRONI78 & 95.1 & 1 & 1.0230 & 0.2024 & 0.6528 & \\
PEDRONI78 & 96.9 & 1 & 1.0166 & 0.1305 & 0.7179 & \\
\hline
\end{tabular}
\end{center}
\end{table*}

\newpage
\begin{table}
{\bf \caption{\label{tab:2-table}}}The data sets comprising the truncated database for $\pi^- p$ elastic scattering, the pion laboratory kinetic
energy $T$ (in MeV), the number of degrees of freedom $(NDF)_j$ for each set, the scale factor $z_j$ which minimises $\chi_j^2$, the values of
$(\chi_j^2)_{min}$ and the p-value for each set.
\vspace{0.2cm}
\begin{center}
\begin{tabular}{|l|lrlll|l|}
\hline
Data set & $T$ & $(NDF)_j$ & $z_j$ & $\chi_j^2$ & $p$ & Comments \\
\hline
FRANK83 & 29.4 & 28 & 0.9828 & 31.1504 & 0.3104 & \\
FRANK83 & 49.5 & 28 & 1.1015 & 29.4325 & 0.3908 & \\
FRANK83 & 69.6 & 27 & 1.0931 & 27.0824 & 0.4594 & \\
FRANK83 & 89.6 & 27 & 0.9467 & 24.7108 & 0.5907 & \\
BRACK86 & 66.8 & 5 & 0.9965 & 14.3569 & 0.0135 & \\
BRACK86 & 86.8 & 5 & 1.0032 & 1.3478 & 0.9299 & \\
BRACK86 & 91.7 & 5 & 0.9964 & 3.0272 & 0.6958 & \\
BRACK86 & 97.9 & 5 & 1.0003 & 5.8335 & 0.3228 & \\
WIEDNER89& 54.3& 18 & 1.1563 & 23.5094 & 0.1718 & one point removed, freely f\mbox{}loated\\
BRACK90 & 30.0 & 5 & 1.0215 & 5.2577 & 0.3853 & \\
BRACK90 & 45.0 & 9 & 1.0541 & 12.2642 & 0.1988 & \\
BRACK95 & 87.5 & 6 & 0.9816 & 10.7547 & 0.0963 & \\
BRACK95 & 98.1 & 7 & 1.0067 & 8.8236 & 0.2656 & one point removed\\
JORAM95 & 32.7 & 4 & 0.9937 & 3.7670 & 0.4385 & \\
JORAM95 & 32.7 & 2 & 0.9533 & 5.6487 & 0.0593 & \\
JORAM95 & 45.1 & 4 & 0.9562 & 12.0551 & 0.0169 & \\
JORAM95 & 45.1 & 3 & 0.9459 & 9.4574 & 0.0238 & \\
JORAM95 & 68.6 & 7 & 1.0841 & 14.8484 & 0.0380 & \\
JORAM95 & 68.6 & 3 & 1.0281 & 2.3391 & 0.5051 & \\
JORAM95 & 32.2 & 20 & 1.0587 & 20.8026 & 0.4088 & \\
JORAM95 & 44.6 & 20 & 0.9421 & 30.5855 & 0.0609 & \\
JANOUSCH97& 43.6 & 1 & 1.0427 & 0.1745 & 0.6762 & \\
JANOUSCH97& 50.3 & 1 & 1.0348 & 0.1418 & 0.7065 & \\
JANOUSCH97& 57.3 & 1 & 1.0830 & 4.5260 & 0.0334 & \\
JANOUSCH97& 64.5 & 1 & 1.0152 & 0.0153 & 0.9015 & \\
JANOUSCH97& 72.0 & 1 & 1.3059 & 4.8803 & 0.0272 & \\
\hline
\end{tabular}
\end{center}
\end{table}

\newpage
\begin{table*}
{\bf Table 2 continued}
\vspace{0.2cm}
\begin{center}
\begin{tabular}{|l|lrlll|l|}
\hline
Data set & $T$ & $(NDF)_j$ & $z_j$ & $\chi_j^2$ & $p$ & Comments \\
\hline
ALDER83 & 98.0 & 6 & 1.0338 & 5.1831 & 0.5206 & \\
SEVIOR89 & 98.0 & 5 & 0.9890 & 1.6659 & 0.8932 & \\
HOFMAN98 & 86.8 & 11 & 1.0015 & 6.0355 & 0.8710 & \\
PATTERSON02 & 57.0 & 10 & 0.9377 & 11.2510 & 0.3383 & \\
PATTERSON02 & 66.9 & 9 & 0.9986 & 4.5388 & 0.8725 & \\
PATTERSON02 & 66.9 & 10 & 0.9502 & 17.0121 & 0.0741 & \\
PATTERSON02 & 87.2 & 11 & 0.9827 & 8.5353 & 0.6647 & \\
PATTERSON02 & 87.2 & 11 & 0.9932 & 5.2523 & 0.9183 & \\
PATTERSON02 & 98.0 & 12 & 0.9964 & 7.0659 & 0.8532 & \\
\hline
\end{tabular}
\end{center}
\end{table*}

\vspace{0.5cm}
\begin{table}
{\bf \caption{\label{tab:3-table}}}The values of the seven parameters of the $\pi N$ model obtained from f\mbox{}its to the combined truncated 
elastic-scattering database, chosen using four dif\mbox{}ferent values of $\mathrm{p}_{min}$ (the signif\mbox{}icance level for rejection of data 
points). The uncertainties correspond to the f\mbox{}it with $\mathrm{p}_{min}=0.0027$. 
\vspace{0.2cm}
\begin{center}
\begin{tabular}{|r|rrrr|r|}
\hline
 & 0.0027 & 0.01 & 0.05 & 0.10 & error \\
\hline
$G_{\sigma}(GeV^{-2})$ & 26.76 & 26.72 & 26.72 & 27.04 & 0.85 \\
$K_{\sigma}$ & 0.011 & 0.009 & 0.013 & 0.020 & 0.036 \\
$G_{\rho}(GeV^{-2})$ & 55.07 & 55.04 & 55.05 & 55.69 & 0.61 \\
$K_{\rho}$ & 0.68 & 0.62 & 0.63 & 0.84 & 0.40 \\
$g_{\pi NN}$ & 12.91 & 12.90 & 12.90 & 12.94 & 0.12 \\
$g_{\pi N \Delta}$ & 29.70 & 29.71 & 29.60 & 29.64 & 0.27 \\
$Z$ & -0.528 & -0.530 & -0.521 & -0.510 & 0.059 \\
\hline
\end{tabular}
\end{center}
\end{table}

\newpage
\begin{table}
{\bf \caption{\label{tab:4-table}}}The correlation matrix for the seven parameters of the $\pi N$ model, for the f\mbox{}it to the combined truncated 
elastic-scattering database corresponding to $\mathrm{p}_{min}=0.0027$.
\vspace{0.2cm}
\begin{center}
\begin{tabular}{|r|rrrrrrr|}
\hline
 & $G_{\sigma}$ & $K_{\sigma}$ & $G_{\rho}$ & $K_{\rho}$ & $g_{\pi NN}$ & $g_{\pi N \Delta}$ &$Z$ \\
\hline
$G_{\sigma}$ & 1.0000 & 0.4762 & -0.0903& -0.0288 & 0.1036 & -0.1606 &-0.1928 \\
$K_{\sigma}$ & 0.4762 & 1.0000 & 0.7539 & 0.8226 & 0.9011 & -0.9343 & 0.7502 \\
$G_{\rho}$ &-0.0903 & 0.7539 & 1.0000 & 0.9051 & 0.9058 & -0.8507 & 0.9002 \\
$K_{\rho}$ &-0.0288 &0.8226 &0.9051 &1.0000 &0.9529 &-0.9290 &0.9510 \\
$g_{\pi NN}$ &0.1036 &0.9011 &0.9058 &0.9529 &1.0000 &-0.9499 &0.9239 \\
$g_{\pi N \Delta}$&-0.1606 &-0.9343 &-0.8507 &-0.9290 &-0.9499 &1.0000 &-0.9031 \\
$Z$ &-0.1928 &0.7502 &0.9002 &0.9510 &0.9239 &-0.9031 &1.0000 \\
\hline
\end{tabular}
\end{center}
\end{table}
\vspace{0.5cm}

\newpage
\begin{table}
{\bf \caption{\label{tab:5-table}}}The values of the six $s$- and $p$-wave em-modif\mbox{}ied hadronic phase shifts (in degrees) from the PSA of the
combined truncated elastic-scattering database.
\vspace{0.2cm}
\begin{center}
\begin{tabular}{|l|llllll|}
\hline
$T(MeV)$ & $\tilde{\delta}_{0+}^{3/2}$ & $\tilde{\delta}_{1-}^{3/2}$ & $\tilde{\delta}_{1+}^{3/2}$ & $\tilde{\delta}_{0+}^{1/2}$ & $\tilde{\delta}_{1-}^{1/2}$ & $\tilde{\delta}_{1+}^{1/2}$ \\
\hline
20 & -2.41(3) & -0.23(0) & 1.28(1) & 4.19(3) & -0.37(1) & -0.16(0)\\
25 & -2.80(4) & -0.31(1) & 1.82(1) & 4.67(3) & -0.49(1) & -0.22(1)\\
30 & -3.19(4) & -0.40(1) & 2.44(2) & 5.11(3) & -0.60(1) & -0.28(1)\\
35 & -3.58(4) & -0.50(1) & 3.13(2) & 5.50(3) & -0.71(2) & -0.34(1)\\
40 & -3.98(4) & -0.61(1) & 3.90(2) & 5.86(3) & -0.81(2) & -0.41(1)\\
45 & -4.37(4) & -0.71(2) & 4.76(2) & 6.19(3) & -0.91(2) & -0.47(1)\\
50 & -4.77(4) & -0.82(2) & 5.70(2) & 6.49(4) & -0.99(3) & -0.54(2)\\
55 & -5.17(4) & -0.94(2) & 6.73(3) & 6.78(4) & -1.07(3) & -0.60(2)\\
60 & -5.57(4) & -1.06(3) & 7.86(3) & 7.04(4) & -1.13(3) & -0.67(2)\\
65 & -5.98(4) & -1.18(3) & 9.10(3) & 7.28(5) & -1.18(4) & -0.73(2)\\
70 & -6.40(4) & -1.30(3) & 10.45(3) & 7.50(5) & -1.21(4) & -0.79(3)\\
75 & -6.82(5) & -1.43(4) & 11.92(3) & 7.71(6) & -1.23(5) & -0.86(3)\\
80 & -7.24(5) & -1.56(4) & 13.53(4) & 7.90(6) & -1.24(5) & -0.92(3)\\
85 & -7.67(6) & -1.69(5) & 15.29(5) & 8.07(7) & -1.22(6) & -0.98(4)\\
90 & -8.10(7) & -1.82(5) & 17.20(6) & 8.23(7) & -1.20(6) & -1.04(4)\\
95 & -8.54(8) & -1.96(6) & 19.29(7) & 8.38(8) & -1.15(7) & -1.10(4)\\
100& -8.98(8) & -2.10(6) & 21.56(9) & 8.51(9) & -1.09(7) & -1.16(5)\\
\hline
\end{tabular}
\end{center}
\end{table}

\clearpage

\clearpage

\clearpage

\end{document}